\documentclass[aps,preprint,unsortedaddress,nofootinbib]{revtex4}%
\usepackage{amssymb}
\usepackage{amsfonts}
\usepackage{amsmath}
\usepackage[dvipsnames]{xcolor}
\usepackage{graphicx}%
\providecommand{\U}[1]{\protect\rule{.1in}{.1in}}
\newcommand{\be}{\begin{equation}}
\newcommand{\ee}{\end{equation}}
\newcommand{\bmul}{\begin{multline}}
\newcommand{\emul}{\end{multline}}
\newcommand{\bea}{\begin{eqnarray}}
\newcommand{\eea}{\end{eqnarray}}

\newcommand{\kk}{\mathbf{k}}
\newcommand{\qq}{\mathbf{q}}

\newcommand{\zero}{\mathbf{0}}

\newcommand{\ii}{\mathrm{i}}

\newcommand{\bb}[1]{\left( #1 \right)}

\newcommand{\bbcro}[1]{\left[ #1 \right]}

\begin{document}
\preprint{ }
\title{Collective excitations of superfluid Fermi gases near the transition temperature}
\author{S. N. Klimin}
\affiliation{{TQC, Universiteit Antwerpen, Universiteitsplein 1, B-2610 Antwerpen, Belgium}}
\author{J. Tempere}
\affiliation{{TQC, Universiteit Antwerpen, Universiteitsplein 1, B-2610 Antwerpen, Belgium}}
\affiliation{{Lyman Laboratory of Physics, Harvard University, USA}}
\author{H. Kurkjian}
\affiliation{{TQC, Universiteit Antwerpen, Universiteitsplein 1, B-2610 Antwerpen, Belgium}}
\affiliation{Laboratoire de Physique Th\'{e}orique, Universit\'{e} de Toulouse, CNRS, UPS, France}

\begin{abstract}

Studying the collective pairing phenomena in a two-component Fermi gas,
we predict the appearance near the transition temperature $T_c$
of a well-resolved collective mode of quadratic dispersion.
The mode is visible both above and below $T_c$ in the system's response to
a driving pairing field. When approaching $T_c$ from below,
the phononic and pair-breaking branches, characteristic of the zero temperature behavior,
reduce to a
relatively low energy-momentum region,
where they are replaced by the quadratically-dispersed
pairing resonance, which thus acts as a precursor of the phase transition.
In the strong-coupling and Bose-Einstein Condensate regime,
this mode is a weakly-damped propagating mode associated
to a Lorentzian resonance. Conversely, in the
BCS limit it is a relaxation mode of pure imaginary eigenenergy.
At large momenta, the resonance disappears when it is
reabsorbed by the lower-edge of the pairing continuum.
At intermediate temperatures between 0 and $T_c$, we unify the newly found
collective phenomena near $T_c$ with the phononic and pair-breaking branches
predicted from previous studies, and we exhaustively classify the roots
of the analytically continued dispersion equation, and show that
they provided a very good summary of the pair spectral functions.

\end{abstract}
\date{\today}
\maketitle

\section{Introduction}

The present theoretical investigation is devoted to oscillation-like
and relaxation-like
collective excitations of atomic Fermi superfluids (see, for review, Ref.
\cite{Strinati2018,Turlapov}) in the crossover between the
Bardeen-Cooper-Schrieffer (BCS) pairing regime and the opposite limit of the
Bose-Einstein condensation (BEC) of molecules. An
increasing interest to collective excitations in condensed Fermi gases has
been recently inspired by experimental achievements
\cite{Bartenstein,Kinast,Altmeyer,Tey,Sidorenkov,Hoinka,Kuhn2020}.
Particularly, in Refs. \cite{Hoinka,Kuhn2020}, the spectral function of the
density response of a Fermi gas has been experimentally investigated at finite
momentum and temperature.

The state-of the art of the theory of collective excitations in atomic Fermi
gases shows that there are still unexplored areas, in particular away from the low-temperature,
low-momentum regime.
Combescot \emph{et al}. \cite{Combescot2006} and Diener \emph{et al}.
\cite{Diener2008} analyzed the dispersion of phononic (Anderson-Bogoliubov)
collective excitations in a wide range of momentum at zero temperature. An
analytic study of the dispersion of phonons
has been performed in Ref. \cite{Kurkjian2016}, still at $T=0$. Ohashi and Griffin
\cite{Ohashi2003} studied the order-parameter response functions at $T\neq0$
and identified a resonance interpreted as a damped phononic collective mode.
Last, the nonzero-temperature phonon lifetime
has been calculated using the perturbation scheme
\cite{Kurkjian2017-2} at low temperature,
and beyond the perturbative regime using the analytic
continuation of the Gaussian pair fluctuation (GPF) propagator \cite{Klimin2019}
at higher temperature, in particular near $T_c$.

Besides phononic modes, superconductors and
Fermi superfluids also support a pair-breaking (sometimes called \textquotedblleft
Higgs\textquotedblright) collective branch. This branch, which is intrinsically related
to the existence of a pair-breaking continuum \cite{Engelbrecht,Andrianov1976} in the quasiparticle spectrum,
has been analytically investigated for atomic Fermi gases in the BCS--BEC crossover at zero
temperature \cite{Kurkjian2019}. At $T\neq0$, Ref. \cite{Scirep} predicted
that a pair-breaking mode very similar to the zero-temperature one
exists as long as the wavelength is much larger than the size of the Cooper
pairs.

Near  the  transition  temperature $T_c$ (where, according to BCS theory, the pair
correlation length $\xi_{\rm pair}$ diverges as $\xi_{\rm pair}\propto|T_c-T|^{-1/2}$, and the order
parameter $\Delta$ vanishes in the ordered phase as $\Delta\propto(T_c-T)^{1/2}$),
the region where phononic and pair-breaking modes exist reduces to energies
$\hbar\omega\lesssim\Delta$ and momenta $q\lesssim1/\xi_{\rm pair}$.
The question of whether collective excitations
characteristic of the onset of a superfluid phase
are still visible near $T_c$ at wavevectors
low or comparable to the Fermi wavevector $k_F$, is thus still open.

Here, we show that the dispersion equation
supports a collective branch of quadratic start at and near $T_c$.
This mode is a weakly-damped propagating mode at strong coupling
(it is even undamped in the BEC regime) and a relaxation mode (of a purely imaginary frequency)
in the BCS limit. It generates a well-resolved resonance in the pair spectral function in
the whole BEC-BCS crossover. Observable also at $T>T_c$ provided
one can drive the formation of pairs in the system
(for example by coupling it to a reservoir of superfluid pairs)
this modes acts as a precursor of the superfluid phase transition.
Above $T_c$, it signals that Cooper pairs injected into the system
subsist longer as the temperature approaches $T_c$,
just like ice subsists longer in liquid water
whose temperature approaches $0^\circ $C.
After its quadratic depart, the resonance disappears
at wavevectors $q\gtrsim k_F$ when it is absorbed by the rising lower edge of the
pairing continuum.

At intermediate temperatures between $0$ and $T_c$, we supplement
existing studies by performing an exhaustive cartography of the roots of the dispersion
equation at all momenta, exploring all possible windows of analytic continuation.
The eigenfrequencies and damping factors are determined here mutually consistently, i.~e., beyond the perturbative
approximation for damping. A key advantage of our exhaustive study
is that all collective excitation branches are brought
together within a unified approach. In particular, we explain
that the newly found collective pairing mode at $T=T_c$ differs in nature
from the phononic and pair-breaking branches: its emergence when
$T\to T_c$ is caused by distinct poles of the analytic continuation.
Finally, we compare the spectral function to its estimate based on the poles
(and associated residues) found in the analytic continuation, finding a very
good agreement between the two.

\section{Method}

We consider here a superfluid Fermi gas with $s$-wave pairing. Both
equilibrium and response properties can be determined from the partition
function of the fermionic system. Within the path integral formalism \cite{deMelo1993,Engelbrecht,Diener2008}, the
partition function is a path integral over Grassmann variables $\left(  \bar{\psi}_{\sigma}%
,\psi_{\sigma}\right)  $, which replace the second quantization operators. The model
fermionic action is given by:%
\begin{equation}
S=\int_{0}^{\beta}d\tau\int d\mathbf{r}\left[  \sum_{\sigma=\uparrow
,\downarrow}\bar{\psi}_{\sigma}\left(  \frac{\partial}{\partial\tau}%
-\frac{\nabla_{\mathbf{r}}^{2}}{2m}-\mu\right)  \psi_{\sigma}+g\bar{\psi
}_{\uparrow}\bar{\psi}_{\downarrow}\psi_{\downarrow}\psi_{\uparrow}\right],
\label{S}%
\end{equation}
where we have set $\hbar=k_B=1$. Here $\beta=1/T$ is the inverse to
temperature, $\mu$ is the chemical potential, and $g<0$ is the bare coupling
strength of the model $s$-wave contact interaction.
This coupling constant is renormalized at fixed the scattering length $a_{s}$ by the relation:
\cite{deMelo1993}:%
\begin{equation}
\frac{1}{g}=\frac{m}{4\pi a_{s}}-\int_{k<k_{c}}\frac{d^{3}k}{\left(
2\pi\right)  ^{3}}\frac{m}{k^{2}}\label{g}%
\end{equation}
where $k_{c}$ is the cutoff momentum. Further on, we appply the limit
$k_{c}\rightarrow\infty$, which leads to the contact interaction constant
$g\rightarrow0$.

The Hubbard-Stratonovich transformation introducing the bosonic pair field
$\left[  \bar{\Psi},\Psi\right]  $ with the subsequent integration over the
fermion fields results in an effective bosonic action
\cite{Diener2008,Klimin2019}. Within the Gaussian pair fluctuation (GPF) approximation,
collective modes for a superfluid Fermi gas appear as
fluctuations of this bosonic action on top of the uniform background saddle-point value $\Delta$ of
the pair field. In the present work, background values of the gap $\Delta$ and
the chemical potential $\mu$ are calculated within the mean-field
approximation. This gives us a qualitatively adequate description of the
collective excitations. For a better quantitative description, a equation of
state beyond the mean-field approximation should be applied but this is beyond
the scope of the present treatment.

We determine the spectra of collective excitations within GPF using the method
of the analytic continuation of the GPF matrix elements through their branch
cuts, as proposed by Nozi\`{e}res \cite{Nozieres}. Because the formalism
remains the same as in our preceding works on collective excitations
\cite{Klimin2019,Kurkjian2019}, the scheme of the calculation is reproduced
here only briefly. The complex eigenfrequencies of collective excitations are
determined as the roots of the determinant of the inverse
GPF propagator,%
\begin{equation}
\det\mathbb{M}_\downarrow\left(  \mathbf{q},z_\textbf{q}\right)=0 \label{det}
\end{equation}
The matrix elements of $\mathbb{M}$,
derived in Refs.~\cite{Engelbrecht,Kurkjian2017-2,Klimin2019}
(see \textit{e.g.} Eqs.~(10) and (11) in \cite{Klimin2019}), have a branch cut
all along the real axis, such that roots of \eqref{det} are found only
in the  matrix $\mathbb{M}_\downarrow$ analytically continued to the lower-half
complex plane.

Whereas $\mathbb{M}$ describes the fluctuations
of the order-parameter in the cartesian basis $(\delta\Delta,\delta\Delta^*)$,
it is often easier to deal with fluctuations in the phase-modulus (or phase-amplitude
by a misuse of language we allow ourselves here) basis $(\delta(\text{arg}\Delta),\delta|\Delta|)$
(this is particularly the case at low-temperature and momentum where phase
and modulus fluctuations are well decoupled). The fluctuation matrix in this basis is
\begin{equation}
\mathbb{Q}=P^\dagger \mathbb{M} P
\end{equation}
where $P$ is the hermitian matrix $\begin{pmatrix} 1&\textrm{i}\\ 1&- \textrm{i} \end{pmatrix}/\sqrt{2}$.
The matrix elements of $\mathbb{Q}$ ($Q_{1,1}  $ and $Q_{2,2} $ correspond to the amplitude
and phase fluctuations, respectively, and $Q_{1,2}  $ describe mixing of amplitude and phase
fluctuations) are given in appendix \ref{app:matrix}.

Strictly speaking, complex poles of Green's functions in a condensed matter
theory can be reliably interpreted as eigenfrequencies and damping factors of
collective excitations or quasiparticles when the damping factors are
relatively small with respect to eigenfrequencies. Nevertheless, they have a heuristic
value even when damping is not small, as long as they bring significant contributions to
the pair field and density spectral functions. Complex poles of the GPF
propagator can reveal the analytic structure and the physical origin of the shape
of the spectral functions, even when this shape is not a simple Lorentzian peak.
In fact, we will show in Sec.~\ref{visibility} that the poles found in the analytic
continuation (together with their associated residues) often constitute
an excellent summary of the spectral function, even when their imaginary part
is comparatively large. This makes the present study relevant for an
explanation of experiments on response properties of cold gases.

\section{Collective mode near the transition temperature}
\label{Tc}

In this section, we concentrate on the collective phenomena
at temperatures close to $T_c$. In Ref.~\cite{Andrianov1976},
it was found that a collective mode whose eigenenergy is purely imaginary
and behaves quadratically in $q$ at low momenta ($q\ll k_F$ with $k_F=(  3\pi
^{2}n)  ^{1/3}$ the Fermi wavevector in terms of the total density $n$)
exists at and near $T_c$ in the BCS limit
($1/k_F a_s\to-\infty$).
Such collective phenomenon where an initial perturbation damps out
without propagating is sometimes called a relaxation mode. On the other side of the crossover,
in the BEC regime ($\mu<0$) at $T=T_c$, Ref.~\cite{Engelbrecht} predicted
a propagating collective mode, with a purely real eigenenergy
but still a quadratic dispersion. Here, we perform
a complete study of this collective mode: we show how it evolves
from a purely imaginary to a purely real mode in the BCS-BEC
crossover, how its eigenenergy varies beyond the long-wavelength
regime and how it is affected by small temperature deviations $|T-T_c|\ll T_c$,
both below and above $T_c$.

\subsection{Effective mass at $T=T_c$ \label{atTc}}

\begin{figure}
\includegraphics{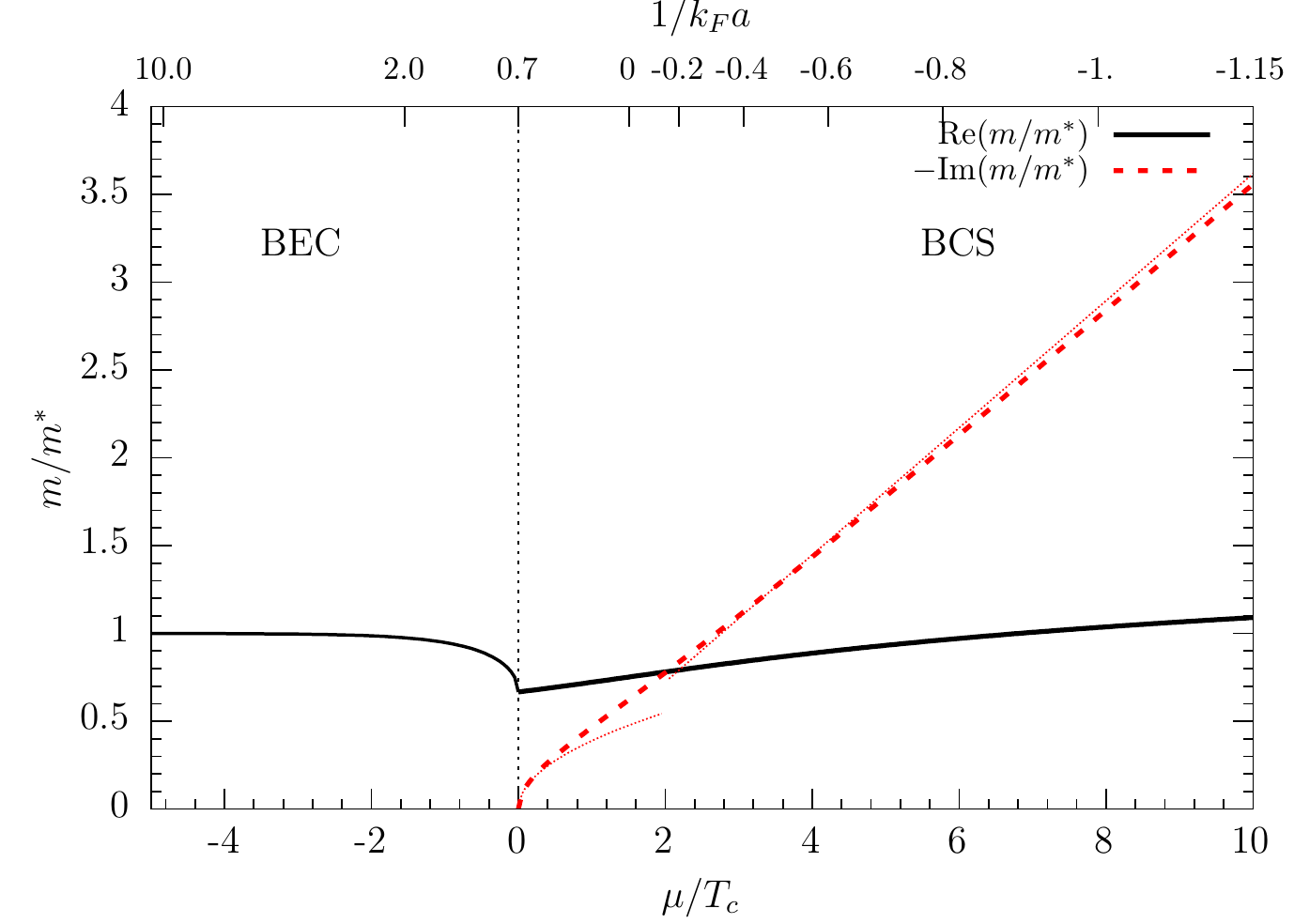}
\caption{\label{fig:effmass} The complex effective mass (real part in black solid line and imaginary part in red dashed line) of the collective mode at $T=T_c$ in the BEC-BCS crossover (values of the dimensionless ratio $\mu/T_c$ on the lower $x$-axis and the dimensionless inverse scattering length $1/k_F a$ and the upper one). The red dotted lines show the limiting behavior of $\textrm{Im}\ m/m^*$ in $\mu_c/T_c\to0^+$ and $\mu_c/T_c\to+\infty$ (see the main text).}
\end{figure}

Exactly at $T_c$, the order-parameter vanishes ($\Delta=0$)
and the fluctuation matrix becomes diagonal
such that the eigenenergy of the collective mode solves simply
\begin{equation}
M_{11}\left(\mathbf{q},z_\mathbf{q}\right)\underset{T=T_c}{=}\sum_{\mathbf{k}}\frac{X(\beta_c\xi_{\mathbf{q}/2+\mathbf{k}})+X(\beta_c\xi_{\mathbf{q}/2-\mathbf{k}})}{2(z_\mathbf{q}-\xi_{\mathbf{q}/2+\mathbf{k}}-\xi_{\mathbf{q}/2-\mathbf{k}})}+ \frac{X( \beta_c\xi_\mathbf{k})}{2\xi_\mathbf{k}}=0 \label{M110}
\end{equation}
with the inverse temperature $\beta_{c}=1/T_{c}$,
$\xi_{\mathbf{k}}=\frac{k^{2}}{2m}-\mu$ the free-fermion energy counted
from the chemical potential $\mu$, and the function $X(\beta\xi)=\tanh(\beta\xi/2)$.
At $T_c$, the interaction regime can be measured by $\mu_c/T_c$ (with $\mu_c\equiv\mu(T_c)$)
as an alternative to $1/k_F a_s$. The BCS and BEC limit
then correspond to $\mu_c/T_c\to+\infty$
and $\mu_c/T_c\to-\infty$ respectively.
In the long wavelength limit ($q\ll k_F$), the only solution of this
equation varies as $q^2$ and is thus characterized by an effective mass $m^*$:
\begin{equation}
z_{\bf q}=\frac{\hbar^2q^2}{4m^*}
\label{zqTc}
\end{equation}
This effective mass, shown on Fig.~\ref{fig:effmass} in the BEC-BCS crossover,
is found by expanding $M_{11}$ at low $q$ and low $z\propto q^2$:
\begin{equation}
\frac{m}{m^*}=\frac{C}{2D}
\end{equation}
with
\begin{align}
C  &  =\int\frac{d\mathbf{k}}{\left(  2\pi\right)  ^{3}}\frac{X\left(\beta_c
\xi_{\mathbf{k}}\right)  -\beta_c\xi_{\mathbf{k}}X^{\prime}\left(  \beta_c\xi_{\mathbf{k}%
}\right)  -\frac{2}{3}\beta_c^2\frac{k^{2}}{2m}\xi_{\mathbf{k}}X^{\prime\prime}\left( \beta_c \xi_{\mathbf{k}%
}\right)  }{8\xi_{\mathbf{k}}^{2}},\label{C}\\
D  &  =D'+\textrm{i}D'' \quad \textrm{with} \quad D'=\mathcal{P}\int\frac{d\mathbf{k}%
}{\left(  2\pi\right)  ^{3}}\frac{X\left( \beta_c \xi_{\mathbf{k}}\right)  }%
{4\xi_{\mathbf{k}}^{2}},
\quad \textrm{and} \quad D''= \frac{\Theta\left(  \mu\right) }{32\pi}\frac{(2m\mu)^{3/2}}{\mu T_c }\label{D}
\end{align}

In the BEC regime ($\mu_c<0$), the effective mass is real because the fermionic
continuum $\xi_{\mathbf{q}/2-\mathbf{k}}+\xi_{\mathbf{q}/2-\mathbf{k}}$
is gapped (it is bounded from below by its value $2|\mu|+\frac{q^2}{4m}$ in $k=0$)
and does not damp the collective mode \cite{deMelo1993}. When going to the BEC
limit ($1/k_F a_s\to+\infty$ or $\mu_c/T_c\to-\infty$), $m^*$ tends to the fermion mass $m$: the collective
mode there is nothing else than the dispersion relation of free bosonic dimers of mass $2m$.
Higher order bosonic effects not captured by our GPF approach
(such as Landau-Beliaev couplings between collective modes) may
provide additional damping channels for the collective mode, as this is the case
in an atomic BECs.

In the BCS regime ($\mu_c>0$), the fermionic continuum does reach
0, the solution $z_\mathbf{q}$ is found in the analytic continuation\footnote{
The analytic continuation is here trivial: if $V$ is the volume of the gas, one has
$M_{11}(\mathbf{q},\omega\pm\textrm{i}0^+)/V=C\frac{q^2}{2m}-(D'\pm\textrm{i} D'') \omega$ and
thus $M_{11,\downarrow}(\mathbf{q},z)/V=C\frac{q^2}{2m}-(D'+\textrm{i} D'') z$.} of $M_{11}$
and the effective mass acquires an imaginary part. We note the remarkable
value at the threshold of the BEC regime $m/m^*=2/3$ when $\mu_c=0$
and the squareroot growth of the damping coefficient $\textrm{Im}\ m/m^*\underset{\mu_c\to0^+}{\sim}
-0.38\sqrt{\mu_c/T_c}$. At unitary ($1/|a_s|=0$), the real and imaginary part are comparable: $m/m^*=0.752-0.622\,\textrm{i}$.
Finally, in the BCS limit ($1/k_F a_s\to-\infty$ or $\mu_c/T_c\to+\infty$), the imaginary part
diverges as $\mu_c/T_c\approx\epsilon_F/T_c$
\begin{equation}
\frac{m}{m^*}\underset{1/k_F a\to-\infty}{\sim}-\textrm{i}\frac{28\zeta(3)\epsilon_F}{3\pi^3 T_c}
\label{limBCS}
\end{equation}
as found
in Ref.~\cite{Andrianov1976}. The imaginary part thus largely dominates over the real
part which diverges only logarithmically.

\subsection{Long wavelength behavior in the vicinity of $T_c$ \label{vicTc}}

\paragraph{Above $T_c$}
Remarkably, the collective mode found here at
$T_c$, still persists in the normal phase $T>T_c$ as a precursor
of the phase transition:
\begin{equation}
z_{\mathbf{q}}=\frac{q^2}{4m^*}-\alpha(T-T_c)
\label{zqaboveTc}%
\end{equation}
This equation is valid for $q^2/2m\approx |T-T_c|\ll k_F^2/2m,T_c$. It introduces
the additional coefficient
\begin{equation}
\alpha=\frac{E}{T_c^2D} \quad\textrm{with}\quad
E    =-\int\frac{d\mathbf{k}}{\left(  2\pi\right)  ^{3}}\frac{X^{\prime
}\left( \beta_c \xi_{\mathbf{k}}\right)  }{2}. %
\end{equation}
In the BCS limit the shift $-\alpha(T-T_c)$ of the eigenenergy
from its $T=T_c$ value is purely imaginary and leads to
the expression $z_\textbf{q}=-\textrm{i}\frac{28\zeta(3)}{3\pi^3}
\frac{\epsilon_F}{k_{\rm B}T_c} \frac{q^2}{2m}-\textrm{i}\frac{8(T-T_c)}{\pi}$
obtained in \cite{Andrianov1976}. Physically, this means
that Cooper pairs injected in the system have a shorter lifetime
when the temperature rises above $T_c$.
Correspondingly, the visibility of the collective mode fades away as one moves
away from the phase transition.
Conversely, in the BEC regime,
the shift $-\alpha(T-T_c)$ is real positive and acts as a gap
of the collective mode. In both BEC and BCS case, this shift ensures
that $M_{11}(q=0,\omega=0)$ does not vanish in the normal phase,
in accordance with the condition of Nozi\`{e}res -- Schmitt-Rink \cite{Nozieres1985}
and Goldstone theorem.

Although exciting the pair spectral function above $T_c$
is not possible using usual density-coupled probes, this can achieved experimentally by
coupling through a tunneling barrier \cite{Klimin2019}
the sampled gas prepared at $T\gtrsim T_c$ to a reservoir of Cooper pairs at $T\ll T_c$,
as was done for superconductors by Carlson-Goldman \cite{Goldman1976} (we note in passing
the formal analogy between Eq.~\eqref{zqaboveTc} and Eq.~(18--20) in \cite{Goldman1976},
although the spectrum of the mode of Carlson-Goldman
is calculated in a charged fermion gas and with impurities limiting the quasiparticle lifetime).

\paragraph{Below $T_c$}

The detailed evolution of collective modes from $T=0$ to $T_c$
is the subject of the an in-depth numerical study in the next section,
but expression \eqref{zqTc} of the collective mode can already be extended
to temperatures slightly below $T_c$. The picture here is complexified
by the presence of a small number of condensed pairs. Near $T_c$ (and as long
as $\mu$ is positive) the typical size of those pairs $\xi_{\rm pair}=\sqrt{\mu/m\Delta^2}$
diverges, eventually becoming much larger than $1/k_{F}$.
This opens a regime $q\lessapprox 1/\xi_{\rm pair}$ where the physics
of collective modes is similar to what exists at zero temperature
(with phononic modes below $2\Delta$ \cite{Klimin2019} and pair-breaking ``Higgs'' modes above \cite{Scirep}).

We focus here on the intermediate regime $1/\xi_{\rm pair} \ll q \ll k_F$.
Let us first remark that ignoring the presence of a gap $\Delta$
and solving the equation $M_{11}(\qq,z_\qq)=0$ (which applies in the normal phase)
leads to an unstable solution $z_\qq$, with a positive imaginary part
$\operatorname{Im} z_\zero\propto \beta-\beta_c$. This proves the instability of the normal
phase below $T_c$.
Instead, taking into account the deviation $M_{11}(\qq,z,\Delta)-M_{11}(\qq,z,0)$
as well as the non-vanishing off-diagonal element $M_{12}$, one obtains
the quadratic equation:
\begin{equation}
\left(  C\frac{q^{2}}{2m}-iD^{\prime\prime}z_{\mathbf{q}}-\left(  \beta
-\beta_{c}\right)  E\right)  ^{2}-\left(  \beta-\beta_{c}\right)  ^{2}%
E^{2}-\left(  D^{\prime}\right)  ^{2}z_{\mathbf{q}}^{2}=0. \label{APeq}%
\end{equation}
Details on the derivation of this equation are given in Appendix \ref{details}.
Away from the BCS limit, the two solutions of this equation are
not physically distinct $z_{\textbf{q},2}=-z_{\textbf{q},1}^*$ with
\be
z_{\textbf{q},1}=-\ii\frac{CD''}{|D|^2}\frac{q^2}{2m}-\ii\frac{|E|D''}{|D|^2} (\beta-\beta_c)+D'\sqrt{C^2\bb{\frac{q^2}{2m}}^2+2C|E|(\beta-\beta_c){\frac{q^2}{2m}}}
\ee
Here, we have neglected in the discriminant of Eq.~\eqref{APeq} terms of order $(\beta-\beta_c)^2 $.
It is worth noting that although this equation is valid only
for $1/2m\xi_{\rm pair}^2\approx\Delta^2/\mu\ll q^2/2m$ (that is for $(\beta-\beta_c)/\beta_c^2\ll q^2/2m$),
it predicts a transition from a phononic low-velocity regime $z_{\textbf{q},1}\propto (\beta-\beta_c)^{1/2}q$
to a quadratic regime $z_{\textbf{q},1}\propto q^2$ when extrapolated
outside its validity regime to $(\beta-\beta_c)q^2/2m\approx1$. However, we will show
in section \ref{numeric} that in the general case
(and unlike what was found in Ref.~\cite{Andrianov1976} for the BCS limit)
the phononic branch appearing in the regime $q\approx 1/\xi_{\rm pair}$
is supported by a distinct collective branch as the one supporting the quadratic
branch $z_\qq$.

In the BCS limit, the coefficient $D'$ vanishes, and we must reincorporate the term $\propto(\beta-\beta_c)^2 $
to the discriminant. We then obtain the two physically distinct solutions of Ref.~\cite{Andrianov1976}:
\bea
z_{\textbf{q},1}&=& -\textrm{i}\frac{28\zeta(3)}{3\pi^3}
\frac{\epsilon_F}{k_{\rm B}T_c} \frac{q^2}{2m}\\
z_{\textbf{q},2}&=&z_{\textbf{q},1}-\textrm{i}\frac{16(T_c-T)}{\pi}
\eea
We note that below $T_c$, irrespective of the interaction regime,
there exists a solution $z_\qq$ which tends to 0 with $q$, in accordance
again with the criterion of Nozières Schmitt-Rink \cite{Nozieres1985}.

\subsection{Pair spectral function at arbitrary momentum}

\paragraph{Pair-response and spectral functions}

To conclude on the observability of the collective mode, we study its
manifestations in pair-field response matrix $1/M(\qq,z)$.
This response matrix quantifies the susceptibility of the system to an
external complex pairing field $\varphi(\mathbf{r},t)^{\ast}\hat{\psi}_{\downarrow
}(\mathbf{r})\hat{\psi}_{\uparrow}(\mathbf{r})$, i.e. its facility to form
pairs. Generally, we expect the collective modes to
manifest themselves as peaks in the spectral functions.
Since the off-diagonal elements of $1/M$ (in the Cartesian basis) vanish in the limit $T\to T_c^-$  (see Appendix A),
we focus here on the diagonal element:
\begin{align}
\chi\left(  \mathbf{q},\omega\right)    & =\frac{1}{\pi}\operatorname{Im}%
\frac{M_{2,2}\left(  \mathbf{q},\omega+\mathrm{i}0^{+}\right)  }%
{\det\mathbb{M}\left(  \mathbf{q},\omega+\mathrm{i}0^{+}\right)  }\underset{T=T_c}{=}
\frac{1}{\pi
}\operatorname{Im}\frac{1}{M_{11}(\mathbf{q},\omega+\mathrm{i}0^{+})}.
\label{sw1}
\end{align}
We also focus on the imaginary part of the response function (the spectral weight),
which quantifies the capacity of the system to absorb energy injected
at frequency $\omega$.

At $T=T_{c}$, the matrix element $M_{1,1}$ can be expressed as the momentum
integral:
\begin{align}
&  M_{1,1}\left(  \mathbf{q},\omega+\mathrm{i}0^{+}\right)  \nonumber\\
&  =\frac{1}{2\pi^{2}}\int_{0}^{\infty}k^{2}dk\left\{  \frac{X\left(
\beta_{c}\xi_{\mathbf{k}}\right)  }{2\xi_{\mathbf{k}}}+\frac{2m}{\beta kq}%
\ln\left(  \frac{\cosh\left(  \frac{\beta_{c}}{2}\left(  \frac{\left(
k+\frac{q}{2}\right)  ^{2}}{2m}-\mu\right)  \right)  }{\cosh\left(
\frac{\beta_{c}}{2}\left(  \frac{\left(  k-\frac{q}{2}\right)  ^{2}}{2m}%
-\mu\right)  \right)  }\right)  \frac{1}{\omega+\mathrm{i}0^{+}-2\xi
_{\mathbf{k}}-\frac{q^{2}}{4m}}\right\}  \label{M11d}%
\end{align}
Note that the expression above is no longer limited to long wavelengths but
applies also to $q\approx\sqrt{2m|\mu|}$. The denominator of the integrand
vanishes (such that $\chi(\omega)$ is nonzero) as soon as $\omega$ is above
the continuum threshold $\omega_{0}(q)=q^{2}/4m-2\mu$. In this interval, the
spectral function can be
expressed analytically and thus easily extended to $\operatorname{Im}\,z<0$. The
analytic continuation of $M_{11}$ through the interval $[\omega(q),+\infty
\lbrack$ of the real axis is then:
\begin{equation}
M_{1,1,\downarrow}\left(  \mathbf{q},z\right)  =M_{1,1}\left(  \mathbf{q}%
,z\right)  -\Theta\left(  -\operatorname{Im}z\right)  \frac{2\mathrm{i}m}%
{4\pi\beta q}\ln\left(  \frac{\cosh\left(  \frac{1}{4}\beta_{c}\left(
z+\frac{q\sqrt{4\mu-q^{2}+2z}}{2m}\right)  \right)  }{\cosh\left(  \frac{1}%
{4}\beta_{c}\left(  z-\frac{q\sqrt{4\mu-q^{2}+2z}}{2m}\right)  \right)
}\right)  .\label{M11da}%
\end{equation}
where $\Theta$ is the Heaviside step function.
This last expression allows us to find numerically the complex roots
$z_{\mathbf{q}}=\omega_{\mathbf{q}}-i\gamma_{\mathbf{q}}$ of \eqref{M110}
beyond the long-wavelength quadratic regime.

\begin{figure}
\includegraphics[width=0.8\textwidth]{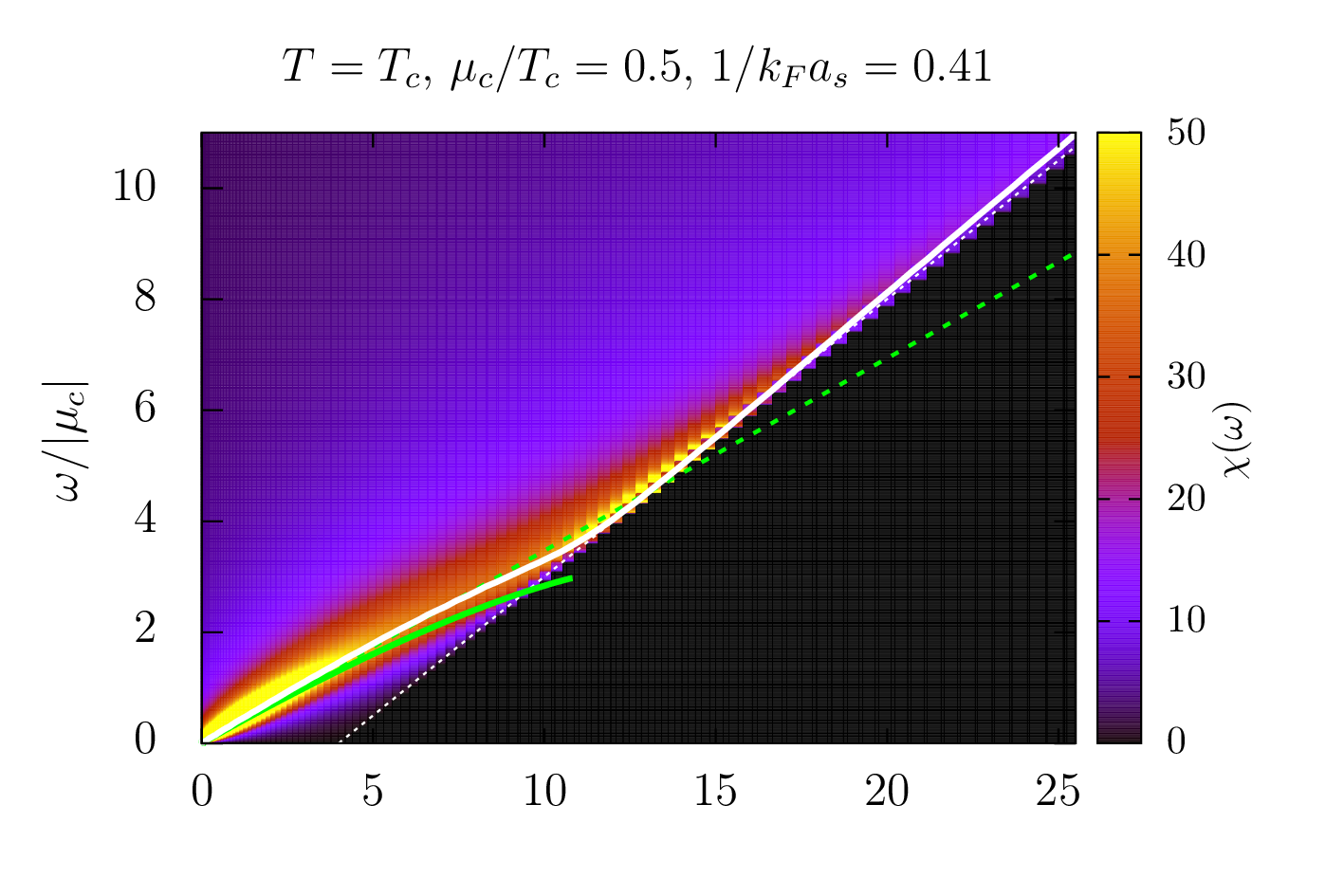}
\includegraphics[width=0.8\textwidth]{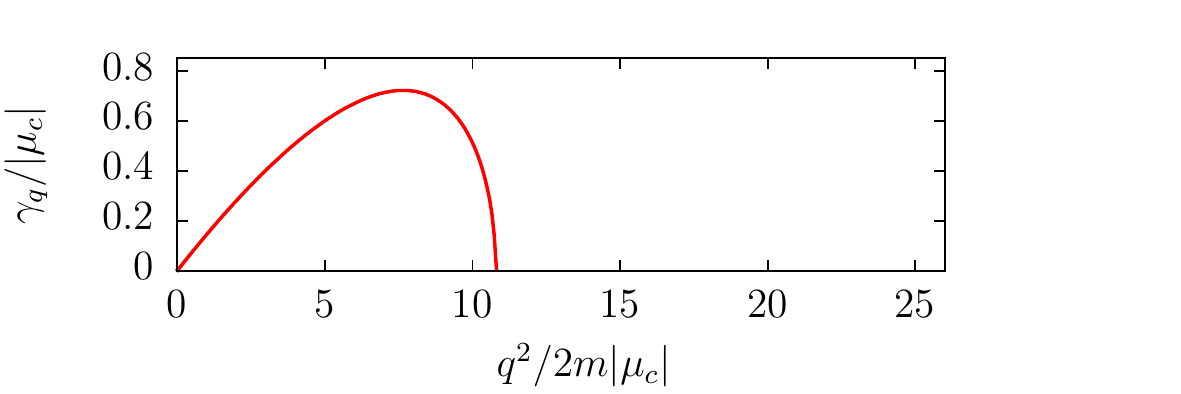}
\caption{\label{fig:strongcoupl}  (Upper panel) The pair spectral
function $\chi(q,\omega+\ii0^+))$ is plotted in colors
as a function of $q^2$ and $\omega$ at $T=T_c$ in the strong-coupling
regime ($1/k_F a\simeq0.41$ and $\mu/T_c=0.5$, still on the BCS side of the crossover).
Here and in subsequent Figs. 3 to 5,
$\chi$ is in units of $|\mu|/\sqrt{2m|\mu|}^3$. Superimposed
to the color plot are the continuum lower edge $\omega_0(q)$ (white dotted line),
the collective mode eigenfrequency $\omega_\qq$ (green solid), its quadratic
low-$q$ expansion (green dashed), and the numerically extracted maximum of the
spectral function (white solid). (Lower panel) The damping rate $\gamma_\qq$
of the collective mode plotted in function of $q^2$ with same $x$ axis as on the upper panel.}
\end{figure}

\begin{figure}
\includegraphics[width=0.8\textwidth]{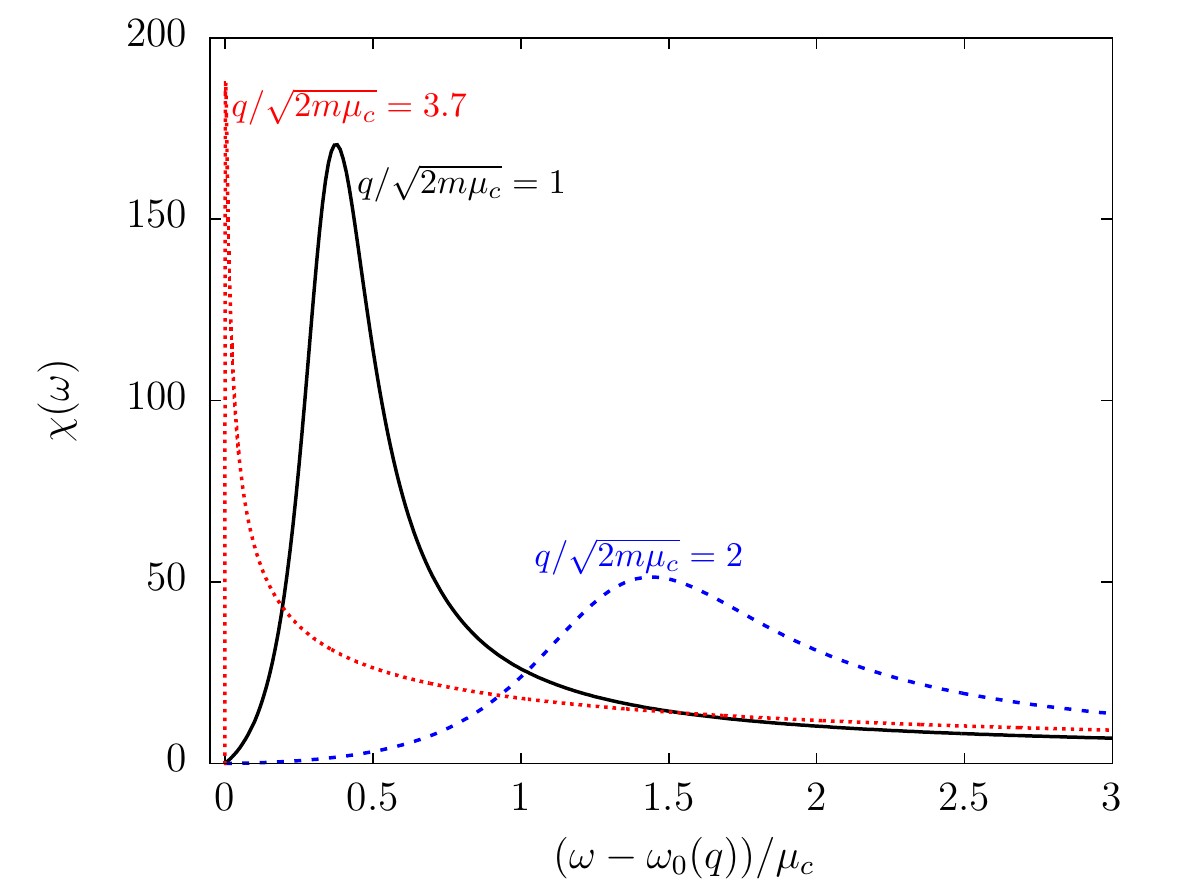}
\caption{\label{fig:coupe} At $T=T_c$ and $1/k_F a\simeq0.41$ (same as on Fig.~\ref{fig:strongcoupl}), the pair spectral function $\chi(q,\omega+\ii0^+))$ is plotted in function of the distance to the continuum edge $\omega-\omega_0(q)$ (with the convention $\omega_0(q)=0$ for $q^2<4m\mu$) at fixed $q=1,2$ and $3.7$ (respectively black solid, blue dashed and red dotted curves).}
\end{figure}

\paragraph{Strong-coupling regime}
On Fig.~\ref{fig:strongcoupl}, we show $\chi$ in function of $q^2$ and $\omega$ as a color plot,
together with the eigenenergy  $\omega_\qq$ [in green (light grey in grayscale)] and damping rate
$\gamma_\qq$ (lower panel) found in the analytic continuation. Notice
that at low $q$ the quadratic variation of the maximum of the spectral function (white curve)
is very well predicted by $\omega_\qq$.
As $q$ increases the peak initially broadens and shifts to higher energy
(see the dashed curve in Fig.~\ref{fig:coupe}).
At some point ($q\simeq3$ on Fig.~\ref{fig:strongcoupl})
$\omega_\qq$ encounters the rising lower-edge of the continuum $\omega_0(q)$ after which
the damping rate falls sharply to $0$. In the spectral function this is associated
with the appearance of a very intense peak pinned at the lower edge of the continuum.
However, unlike at low $q$, this peak of $\chi$ has a large skewness,
with a sharp lower edge and a broad upper tail (dotted curve on Fig.~\ref{fig:coupe}).
It cannot be directly related to a pole in the analytic continuation and thus interpreted
as a collective mode\footnote{We note that a solution of \eqref{det} still exists at $q>3.29$
where the red and green solid lines of Fig.~\ref{fig:strongcoupl} stop. Having a negative
damping rate $\gamma_\qq<0$, this solution is not in the ``physical sector'' (defined
in the introduction of section \ref{numeric}) and can hardly be interpreted as a collective mode.}.

\begin{figure}
\includegraphics[width=0.85\textwidth]{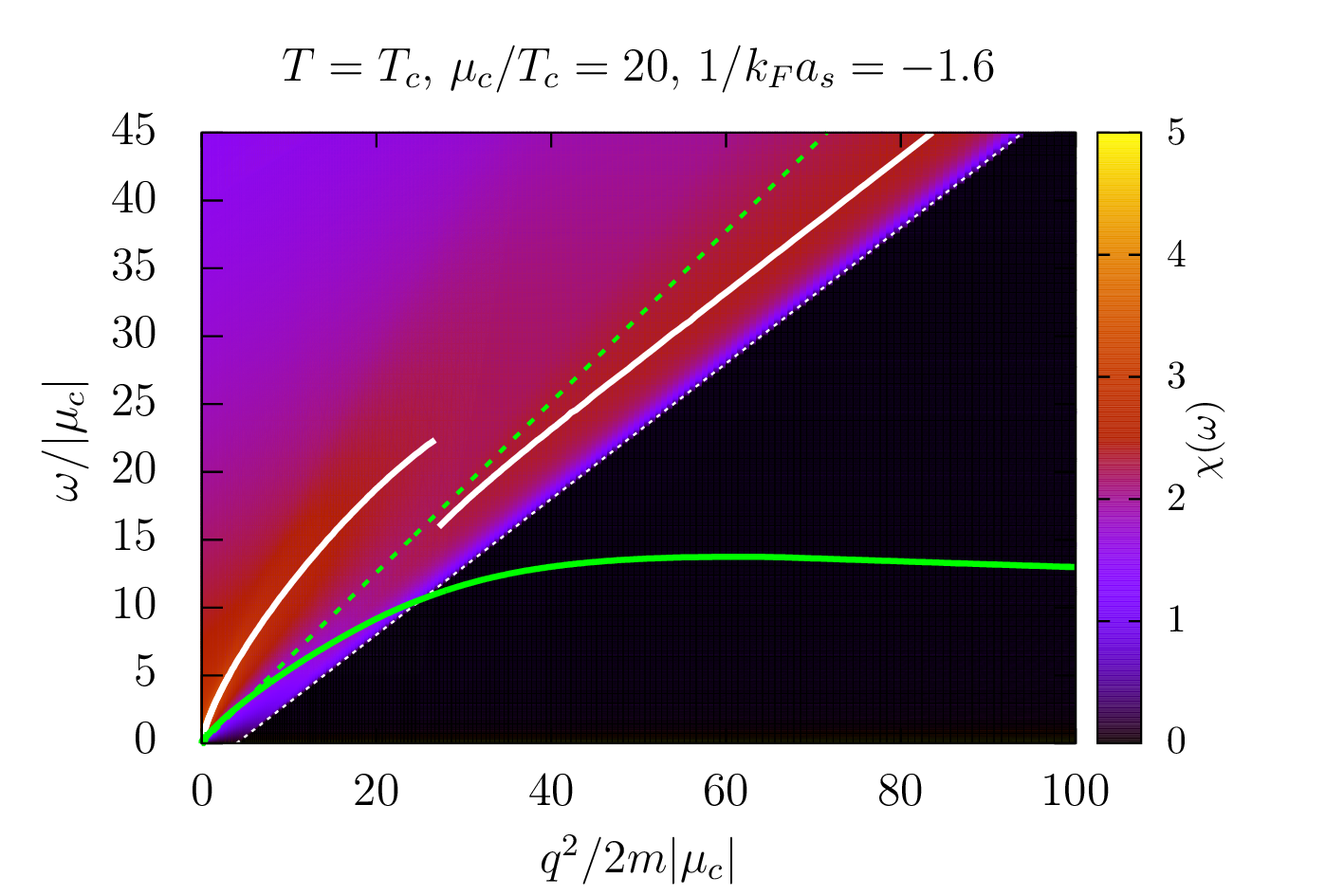}\\
\includegraphics[width=0.85\textwidth]{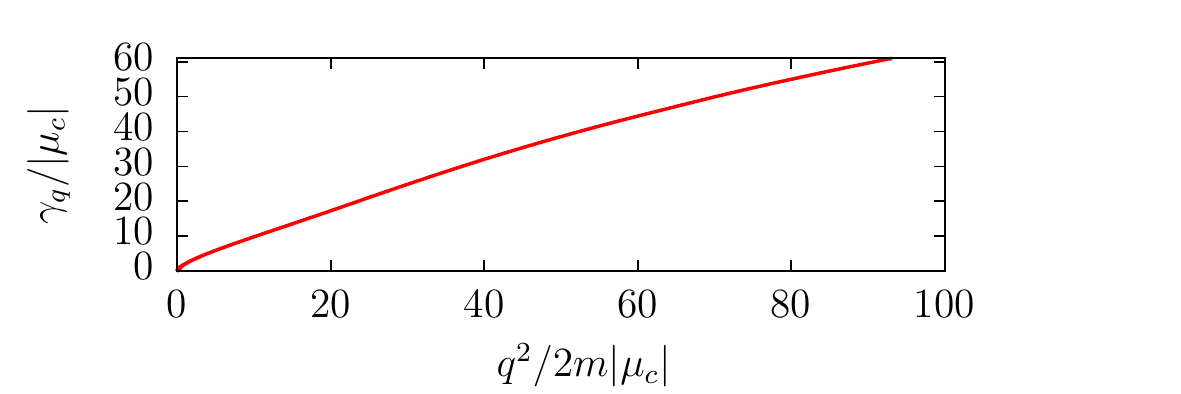}
\caption{\label{fig:BCS} (Upper panel) The pair spectral function $\chi(q,\omega+\ii0^+))$ in the BCS regime ($1/k_F a\simeq-1.6$). The curves superimposed to the color plot are the same as on Fig.~\ref{fig:strongcoupl}. Note the discontinuity of the location of the spectral function maximum, when the sharp peak appears near the continuum edge. (Lower panel) The damping rate $\gamma_\qq$.}
\end{figure}

\paragraph{BCS regime}
When going to the BCS regime (Fig.~\ref{fig:BCS}), the low-$q$ damping of the collective mode
increases (as prescribed by \eqref{limBCS}) such that the resonance fades out quicker. The skewness
of the resonance (which no longer fits to a Lorentzian function \cite{Kurkjian2019}) is also larger, such
that the peak maximum is displaced from $\omega_\qq$. The sharp peak which appears at high $q$ near the continuum
lower-edge is also much less intense than in the strong-coupling regime.

\begin{figure}
\includegraphics[width=0.85\textwidth]{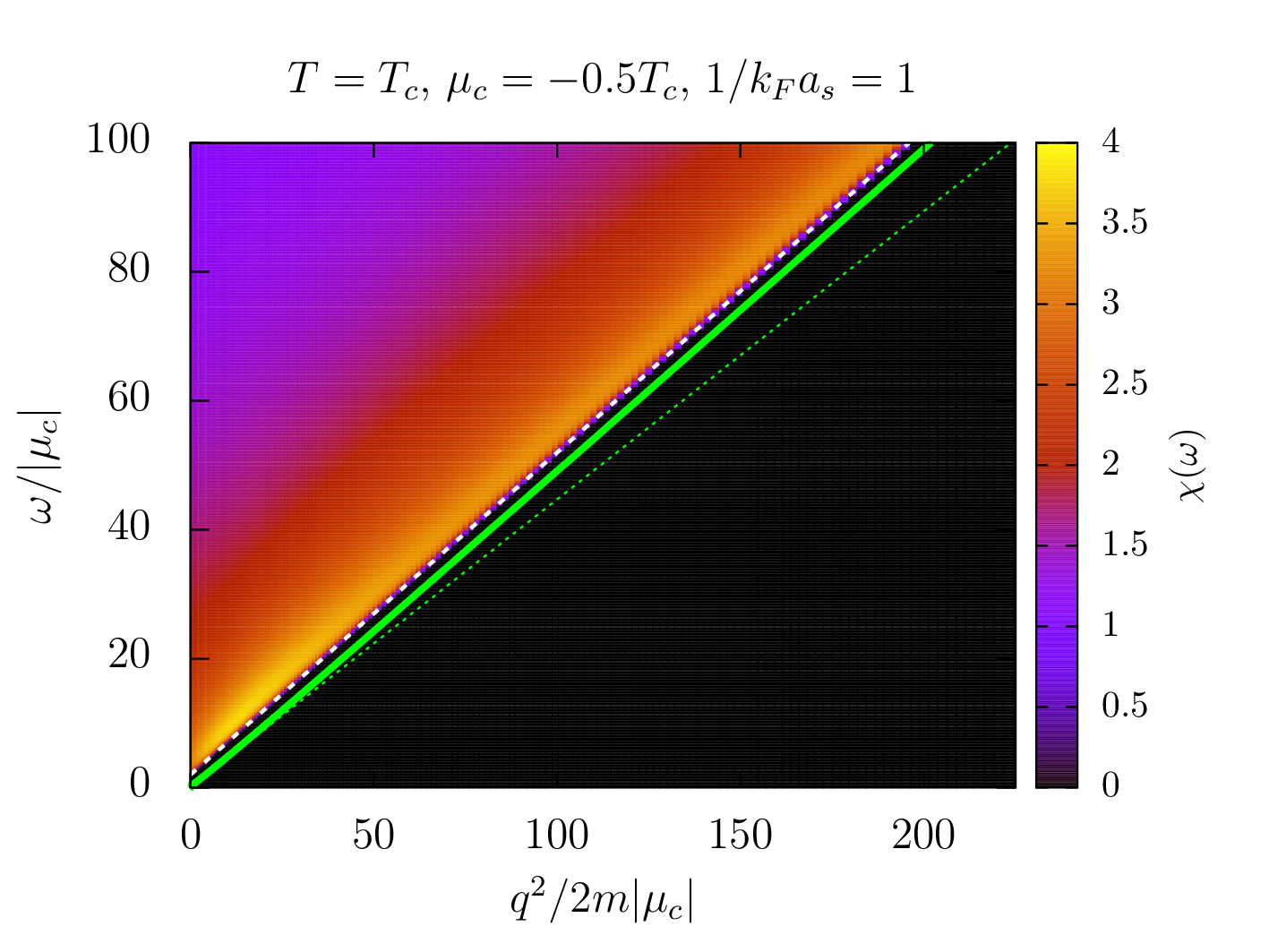}
\caption{\label{fig:BEC} The pair spectral function $\chi(q,\omega+\ii0^+))$ is plotted in colors
as a function of $q^2$ and $\omega$ at $T=T_c$ and in the BEC regime ($1/k_Fa_s\simeq1.0$ and $\mu_c/T_c=-0.5$).
The eigenfrequency of the collective mode $\omega_\qq$ (green solid line), its low-$q$
expansion (dotted green) and the lower-edge of the continuum $\omega_{0}(q)$ (dashed white)
are superimposed to the color plot.}
\end{figure}

\paragraph{BEC regime}
Fig.~\ref{fig:BEC} shows the spectral function in the BEC regime
($1/k_Fa\simeq1.0$ and $\mu/T_c=-0.5$). In this regime a Dirac
peak corresponding to an undamped collective mode (green solid line)
exists below the lower-edge of the continuum $\omega_{0}(q)$ (dashed white line).
The spectral function in the continuum is smooth with a shallow maximum near the continuum edge.
At large $q$, the eigenenergy $\omega_\qq$ tends from below to the threshold energy $\omega_{\rm 0}(q)$.

\section{Collective excitations below the transition temperature}
\label{numeric}
In this section we perform a complete study of the solutions
of the dispersion equation \eqref{det}, at arbitrary momentum,
in the whole temperature range $0<T<T_c$,
and both below (Sec.~\ref{phononic}) and above (Sec.~\ref{continuum}) the pair-breaking continuum threshold.
Generally, collective excitations can be reliably attributed to known
types, such as Anderson-Bogoliubov (phase), pair-breaking (amplitude) modes,
or the pairing mode found at $T_c$ only in limiting cases (such as
$q\to0$, $T\to0$ or $T\to T_c$). At temperatures away from $0$ and $T_c$,
the spectral function may have a non-trivial structure with
more than one maximum, and correspondingly
two (or more) interfering poles \cite{Klimin2019} in the analytic continuation.
This motivates an exhaustive cartography
of all poles of the analytic continuation, as a basis of a rigorous classification
of all collective phenomena.

\begin{figure}
\includegraphics{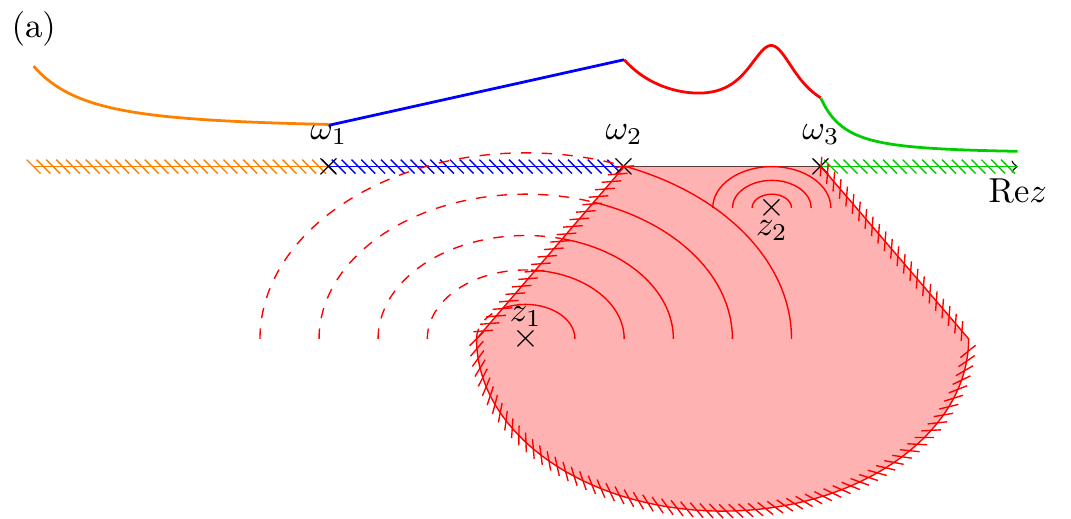}

\includegraphics{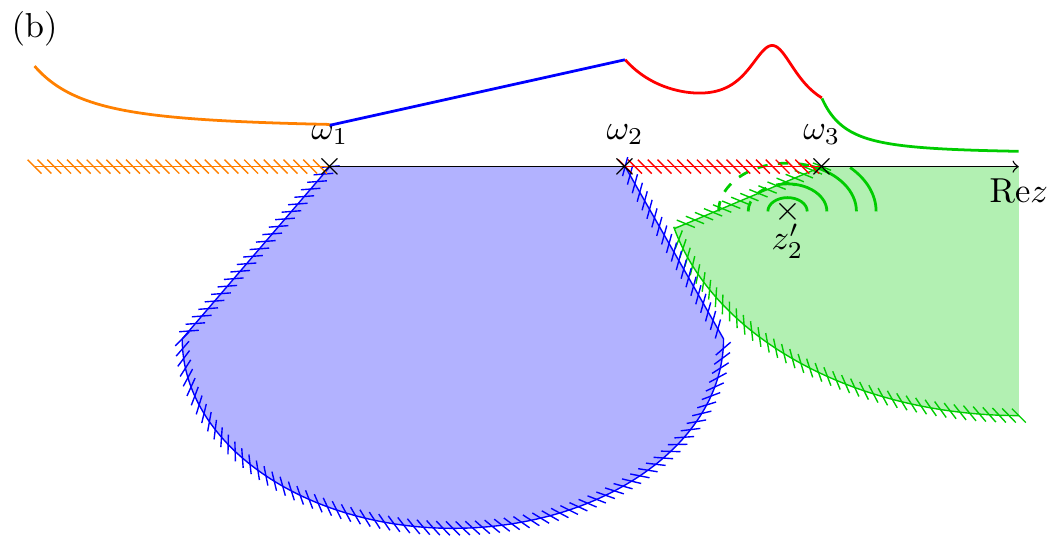}
\caption{{\textbf{Analytic continuations of a response function with multiple angular points}. The analytic continuation of $\chi$ from the interval $[\omega_2,\omega_3]$ to the lower-half-complex plane (red shaded area on panel (a)) has two poles in $z_1$ and $z_2$. The first pole has a real part comprised between $\omega_1$ and $\omega_2$ and gives rises to a resonance in $\chi(\omega)$ (visible on the solid curve to the right of the interval $[\omega_2,\omega_3]$). This pole has a counter-part $z_1'$ in the analytic continuation through $[\omega_3,+\infty[$ (green shaded area on the right of panel (b)), such that the resonance peak is only slightly broken in $\omega_3$ and appears to extend beyond it. This contrasts with the pole $z_2$, which has no counter-part in the continuation through $[\omega_1,\omega_2]$ (blue shaded area on the left of (b)). The ``resonance'' then abruptly terminates in $\omega_2^+$, such that only its upper tail is visible in $\chi(\omega)$ (to the left of the interval $[\omega_2,\omega_3]$). }\label{fig:schema}}
\end{figure}

The analytic continuation of the GPF propagator in the general case raises subtle questions,
which require a careful analysis. As soon as several angular points appear on the
real axis, they determine several analyticity intervals, which open distinct windows for the
analytic continuation (see Fig.~\ref{fig:schema}). A straightforward and \textquotedblleft
naive\textquotedblright\ way to determine complex roots of Eq. (\ref{det})
would be to choose a piecewise rule for the analytic continuation, where the
branch cut on the real axis is converted to vertical branches attached to each branching
point and extending to the lower-half complex plane.
However, other continuation schemes are possible, as described in Refs.
\cite{Klimin2019,Kurkjian2019,Scirep,Castin}. We can in particular extend the analytic
continuation through a chosen window to the entire lower half of the complex plane
(see, e.~g., Fig. 2 of Ref. \cite{Kurkjian2019}). As a result, each window can
provide poles whose real part is outside the continued interval of analyticity.

Mathematically, the analytic continuation of $\mathbb{M}$
is defined in infinite layers of Riemann sheets obtained
by winding around the branching points between which the continuation
is performed. Ref.~\cite{Andrianov1976} showed for instance
that a monodromic infinity of poles are obtained at $T=0$ by
winding around the point $\omega=2\Delta$. This being said,
poles lying far away from the original branch cut
have a smaller impact on the response function, and
thus little physical significance. For this reason, we restrict
our exploration of the analytic continuation to the ``physical sector'' defined as the
fourth quadrant ($\operatorname{Re}z>0$ and $\operatorname{Im}z<0$) of the first Riemann sheet\footnote{Without
loss of generality, we study the spectral functions only at $\omega>0$.}.
However one should keep in mind that solutions initially belonging to ``unphysical'' sectors
of the analytic continuation (such as \textit{e.g.} the quadrant $\operatorname{Re}z<0$, $\operatorname{Im}z<0$)
may eventually enter the physical sector \cite{Castin,Klimin2019} as $q$ and $T$ vary.
Or, vice-versa, poles of the physical
sector may eventually leave it.


\subsection{Angular points and intervals for the analytic continuation}

The analytic continuation is performed using the standard scheme. Consider a
function $F$ of the complex variable $z$ having a branch cut on the real axis
$z=\omega$ and introduce the associated spectral density,%
\begin{equation}
\rho_{F}\left(  \omega\right)  =-\lim_{\delta\rightarrow0}\frac{F\left(
\omega+i\delta\right)  -F\left(  \omega-i\delta\right)  }{2\pi i}. \label{rho}%
\end{equation}
The spectral density $\rho_{F}\left(  \omega\right)  $ is in general analytic
on the real axis except at most on a finite number of points. It can thus be
analytically continued from any chosen interval between these points to the
lower complex half-plane. The analytic continuation $F^{\left(  I\right)
}\left(  z\right)  $ of $F\left(  z\right)  $ from upper to lower complex
half-plane and through the interval $I\subset\mathbb{R}$ where $\rho_{F}$ is
analytic then reads:%
\begin{equation}
F^{\left(  I\right)  }\left(  z\right)  =\left\{
\begin{array}
[c]{cc}%
F\left(  z\right)  , & \operatorname{Im}z>0,\\
F\left(  z\right)  -2\pi i\rho_{F}^{(I)}\left(  z\right)  , &
\operatorname{Im}z<0,
\end{array}
\right.  \label{ancont}%
\end{equation}
where $z\rightarrow\rho_{F}^{(I)}\left(  z\right)  $ is the analytic
continuation of $\rho_{F}\left(  \omega\right)  $ from the interval $I$ to the
lower complex half-plane.


The angular points of the spectral density mark a change
in the configuration (usually in the number of connected components)
of the resonant wavevectors for one of the two resonance conditions:
\begin{equation}
E_{\mathbf{k}-\frac{\mathbf{q}}{2}}+E_{\mathbf{k}+\frac{\mathbf{q}}{2}}%
=\omega,\quad\left\vert E_{\mathbf{k}-\frac{\mathbf{q}}{2}}-E_{\mathbf{k}%
+\frac{\mathbf{q}}{2}}\right\vert =\omega \label{eq1}%
\end{equation}
where $E_{\mathbf{k}}=\sqrt{\xi_{\mathbf{k}}^{2}+\Delta^{2}}$ is the BCS
excitation energy and $\xi_{\mathbf{k}}=k^{2}-\mu$ the free-fermion energy
(in this section we set $\hbar=k_{\rm B}=2m=1$, which is equivalent to working in
Fermi units $\epsilon_F$ and $k_F$ respectively for energies and wavevectors),
and we assume $\omega>0$ without loss of generality.
The angular points obtained from the first resonance condition are essential for the analytic
continuation both for zero and nonzero temperatures. They affect the
\textquotedblleft particle-particle\textquotedblright\ terms in the matrix
elements. The angular points obtained from the second resonance condition affect the
\textquotedblleft particle-hole\textquotedblright\ terms and must be taken
into account only when $T\neq0$.

As described in Ref. \cite{Kurkjian2019}, there exist three frequencies
corresponding to angular points in the zero temperature case. The frequency
$\omega_{1}$ is the boundary of the pair-breaking continuum,%
\begin{equation}
\omega_{1}=\left\{
\begin{array}
[c]{cc}%
2\Delta, & \mu-q^{2}/4\geq0,\\
2\sqrt{\left(  \mu-q^{2}/4\right)  ^{2}+\Delta^{2}}, & \mu-q^{2}/4<0.
\end{array}
\right.  \label{w1}%
\end{equation}
The frequency $\omega_{3}=2\sqrt{\left(  \mu-q^{2}/4\right)  ^{2}+\Delta^{2}}$
is the energy of the BCS pair $E_{\mathbf{k}-\frac{\mathbf{q}}{2}%
}+E_{\mathbf{k}+\frac{\mathbf{q}}{2}}$ at $k=0$. For $\mu-q^{2}/4<0$, these frequencies coincide
$\omega_{1}=\omega_{3}$.

\begin{figure}[tbh]%
\centering
\includegraphics[
height=2.7657in,
width=3.5795in
]%
{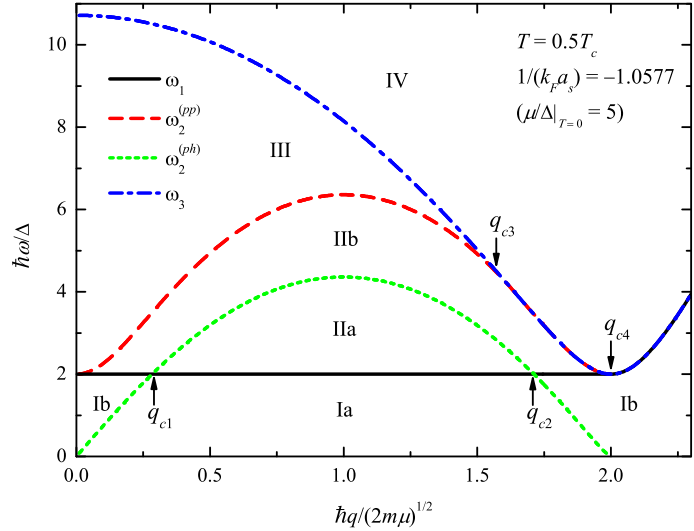}%
\caption{Angular-point frequencies for the analytic continuation of the GPF
matrix elements for $1/k_{F}a_{s}\approx-1.0577$ and $T=0.5T_{c}$. The areas
between curves determine intervals for the analytic continuation as described
in the text. The arrows show values of momentum $q_{c1},\ldots q_{c4}$ at
which different angular-point frequencies coincide.}%
\label{fig:AngPoints}%
\end{figure}

The other angular point frequencies correspond to local minima/maxima of the
energies $E_{\mathbf{k}-\frac{\mathbf{q}}{2}}\pm E_{\mathbf{k}+\frac
{\mathbf{q}}{2}}$ at $\cos\theta_{\mathbf{k},\mathbf{q}}=\pm1$, where
$\theta_{\mathbf{k},\mathbf{q}}$ is the angle between $\mathbf{k}$ and
$\mathbf{q}$.
They are provided by solutions of the equations
\begin{align}
\frac{\partial\left(  E_{{k}-\frac{{q}}{2}}+E_{{k}%
+\frac{{q}}{2}}\right)  }{\partial k}  &  =0,\label{pp}\\
\frac{\partial\left(  E_{{k}-\frac{{q}}{2}}-E_{{k}%
+\frac{{q}}{2}}\right)  }{\partial k}  &  =0, \label{ph}%
\end{align}
which lead to the equation for $\varepsilon\equiv k^{2}$, unique for for both
particle-particle and particle-hole angular points,%
\begin{align}
&  256\varepsilon^{4}-256\left(  q^{2}+4\mu\right)  \varepsilon^{3}+32\left(
8q^{2}\mu+3q^{4}+48\mu^{2}+24\Delta^{2}\right)  \varepsilon^{2}\nonumber\\
&  +16\left(  8\Delta^{2}\left(  5q^{2}-8\mu\right)  -\left(  q^{2}%
+4\mu\right)  \left(  q^{2}-4\mu\right)  ^{2}\right)  \varepsilon\nonumber\\
&  +\left(  q^{2}-4\mu\right)  \left(  \left(  q^{2}-4\mu\right)
^{3}+16\Delta^{2}\left(  3q^{2}-4\mu\right)  \right)  =0. \label{bound2}%
\end{align}
This equation provides up to four (restricted by the additional condition
$\varepsilon>0$) frequencies. They can be classified as $\left(  \omega
_{2s,1},\omega_{2s,2}\right)  $ which satisfy (\ref{pp}), and $\left(
\omega_{2a,1},\omega_{2a,2}\right)  $ which satisfy (\ref{ph}). The ordering
of the frequencies is chosen in such a way that $\omega_{2s,1}<\omega_{2s,2}$
and $\omega_{2a,1}>\omega_{2a,2}$. The chosen ordering for $\left(
\omega_{2s,1},\omega_{2s,2}\right)  $ coincides with the selection of the root
for $\omega_{2}$ in Ref. \cite{Castin}. In this classification, $\omega_{2}$
of the preceding work \cite{Kurkjian2019} coincides with $\omega_{2s,1}$. An
example of these real solutions, together with $\omega_{1}$ and $\omega_{3}$,
is shown in Fig. \ref{fig:AngPoints} for $\left.  \mu/\Delta\right\vert
_{T=0}=5$, that corresponds to the inverse scattering length $1/k_{F}%
a_{s}\approx-1.0577$. The temperature is here $T=0.5T_{c}$.%

After selecting physically reasonable roots from the aforesaid four ones, we
find that two frequencies $\omega_{2s,1}\equiv\omega_{2}^{\left(  pp\right)
}$ and $\omega_{2a,1}\equiv\omega_{2}^{\left(  ph\right)  }$ correspond,
respectively, to the particle-particle and particle-hole angular points. The
particle-particle angular point frequency $\omega_{2}^{\left(  pp\right)  }$
is important both at zero and nonzero temperatures and is described in Ref.
\cite{Kurkjian2019}. The particle-hole angular point frequency $\omega
_{2}^{\left(  ph\right)  }$ contributes only at $T\neq0$. Particularly, the
frequency $\omega_{2}^{\left(  ph\right)  }$ behaves linearly at small
momentum, thus affecting phononic modes. In the small-momentum limit,
the particle-hole angular point frequency asymptotically tends to $c_{b}q$,
where $c_{b}$ is the boundary sound velocity determined in Ref.
\cite{Klimin2019} and corresponding to the opening/closing of a decay channel
in the part of the BCS excitation branch at $k<\sqrt{2m\mu}$.

For any given $q$, intervals between different angular-point frequencies
determine windows for the analytic continuation. Consequently, they are
described by the areas in Fig. \ref{fig:AngPoints} between different curves.
The classification of the windows in the figure, performed by Roman numbers,
extends their zero-temperature classification of Ref. \cite{Castin} to nonzero
temperatures due to the appearance of the particle-hole angular point
frequency $\omega_{2}^{\left(  ph\right)  }$. Below the pair-breaking
continuum, this leads to the subdivision of the interval I to two intervals
Ia and Ib, respectively below and above $\omega_{2}^{\left(  ph\right)  }%
$. We introduce here several critical values of $q$ at which different
angular-point frequencies cross or touch each other. The particle-hole angular
point frequency $\omega_{2}^{\left(  ph\right)  }$ may cross the pair-breaking
continuum edge $\omega_{1}$, as shown in the figure. In this case, also the
interval II is subdivided to two intervals IIa and IIb, as shown in the
figure, and we denote $q_{c1}$ and $q_{c2}$ the crossing points. Therefore at
nonzero temperatures the angular point $\omega_{2}^{\left(  ph\right)  }$ is
important for both phononic and pair-breaking collective excitations. This is
a non-trivial difference between the zero-temperature and nonzero-temperature
cases. As can be seen from the figure, the particle-particle angular point
frequency $\omega_{2}^{\left(  pp\right)  }$ exists for $q$ smaller than a
critical value $q\equiv q_{c3}$, above which it coincides with $\omega_{3}$.
Finally, the pair-breaking continuum edge and $\omega_{3}$
become equal to each other at $q>q_{c4}$, where $q_{c4}\equiv2\sqrt{2m\mu}$.
In the following subsection, we index the obtained solutions of Eq.~\eqref{det}
according to the classification of intervals described in Fig.
\ref{fig:AngPoints}: for exemple $\omega_{\rm IIb}^{(i)}$ will be the $i$-th
root of $\text{det} \mathbb{M}$ found in the analytic continuation through
window IIb.

When analytic continuations through two
adjacent intervals separated by an angular point have drastically
different analytic structures, the shape of the spectral function
abruptly changes at the angular point, with for instance the sudden termination
of a resonance peak (see schematically the blue and red intervals in Fig.~\ref{fig:schema}).
At low temperature, the lower edge $\omega_1$ of the pair-breaking continuum
is such a sharp angular point dividing the frequencies into low- and high-energy regions
with much different physics. On the contrary, adjacent intervals
may yield similar analytic structures, with poles in particular
lying close to one another (as was the case in Ref.~\cite{Klimin2019} for phononic
poles computed from above or below $\omega_2^{\rm (ph)}$,
see schematically the green and red intervals in Fig.~\ref{fig:schema}).
In this case, the spectral function, despite a small kink at the angular
point maintains an overall similar shape on both sides of it, and the
poles in different windows can be attributed to the same physical phenomenon.
In what follows, we consider poles belonging to different continuation windows to be physically equivalent
when their energy separation is smaller than their inverse lifetime,
making them nearly indistinguishable on the real axis.

The equivocality of the complex eigenfrequencies shows the limits
to the concept of quasiparticle. In a system of interacting particle,
this concept is an \emph{approximation} used to grasp the
most stringent features of a continuous spectrum.
Whereas interpreting a single Lorentzian resonance
in terms of a complex pole is straighforward, even
for broad resonances, doing so in case of multiple
resonances, or of an asymmetric, non Lorentzian,
resonance is less obvious. In these cases, the knowledge
of the analytic continuation can help distinguish between
peaks that can be related to complex poles (although
they may be distorted by the continuum background \cite{Kurkjian2019}),
and thus interpreted as resonances,
and peaks which only correspond to a continuum edge.

\subsection{Phononic-like collective excitations}
\label{phononic}

Frequencies and damping factors for phononic collective excitations in the
long-wavelength limit at nonzero temperatures have been calculated in Ref.
\cite{Klimin2019} using the analytic continuation of the inverse GPF propagator
expanded at small $q$.  Contrary to a naive
expectation of two branches for a complex field with a modulus and a phase,
the spectrum of collective excitations at nonzero temperature can contain more than two branches, due
to the additional degrees of freedom provided by the normal component.
As shown in Ref. \cite{Klimin2019}, the spectrum contains in particular
two phononic branches, reminiscent of the first and second sound modes
of the hydrodynamic theory of superfluids. Here, we extend the study of
those two branches to finite momentum. We recall that at zero temperature
the only phononic branch (identified as hydrodynamic first sound)
tends to the pair-breaking continuum threshold, either at a
finite wavevector $q_{\rm sup}$ in the BCS regime,
or asymptotically in $q=+\infty$ in the BEC regime.
At temperatures low compared to $T_c$  \cite{Kurkjian2016},
and in fact in a relatively wide temperature range below $T_c$,
the dispersion shows qualitatively the same features as in the zero-temperature case,
and the damping rate behave as in Ref.~\cite{Kurkjian2016} (falling off to 0
both when $q\to0$ and when the branch approach the pair-breaking continuum
threshold).

We are interested here rather in the regime
of $T$ close to $T_c$ where the second branch enters the physical sector
(coming from the third quadrant of the complex plane).
We identify two remarkable phenomena in the
finite $q$ behavior of the two branches. First, the mechanism
which confines the phononic branch below the pair-breaking
continuum at $T=0$ is lifted here because of the presence
of the particle-hole scattering channel. One of the two branches
thus enters the pair-breaking continuum at large $q$
(while the other leaves the physical sector). Second, in the
BCS regime, the two branches exchange their high-$q$ behavior
at some remarkable temperature: while at low $T$, the ``first branch''
(the one which evolves from the $T=0$ first sound)
reaches the continuum, at temperatures close to $T_c$,
it is instead the second branch. Separating the two scenarios
is an exceptional temperature $T_{ex}$ where the two branches exactly meet
(both in real and imaginary part) at an exceptional momentum $q_{ex}$.
We note that this section adopts a rather theoretical perspective
on the collective modes, as the phenomena we describe
are hardly observable on the spectral functions.

\begin{figure}[ptbh]%
\centering
\includegraphics[
height=4.1978in,
width=6.0416in
]%
{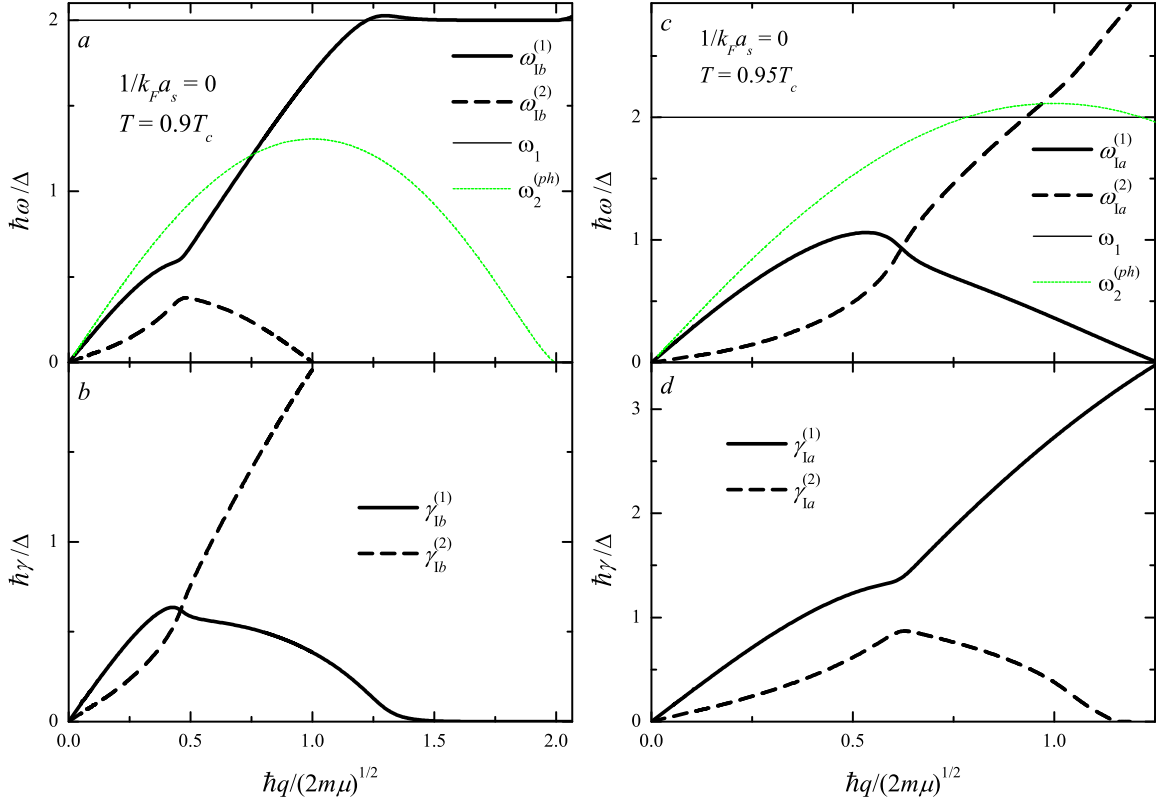}%
\caption{Momentum dispersion of the frequency (\emph{a, c}) and damping
(\emph{b, d}) (measured in units of $\Delta$) of collective modes obtained using the analytic continuation in
frequency intervals below the pair-breaking continuum at $1/\left(  k_{F}%
a_{s}\right)  =0$ with $T/T_{c}=0.9$ (\emph{a, b}) and $T/T_{c}=0.95$
(\emph{c, d}). The classification of intervals follows the scheme described in
Fig. \ref{fig:AngPoints}. Heavy solid and dashed curves show the solutions
corresponding, respectively, to the first and second pole of the GPF
propagator. Thin curves indicate the angular-point frequencies $\omega_{1}$
(the pair-breaking continuum edge) and $\omega_{2}^{\left(  ph\right)  }$ (the
particle-hole angular point frequency).}%
\label{fig:AB1}%
\end{figure}

In Fig. \ref{fig:AB1}, we consider\footnote{We note that
employing window Ia or Ib for the analytic continuation results \textit{a priori}
in distinct solutions. However, the energy mismatch introduced by
the change of window remains small (in particular smaller than the
imaginary part of the poles) at all momenta. For this reason,
we show results only for the window which seems best adapted to follow the dispersion
of the branch. In the case when $\max_q \omega_{2}^{\left(  ph\right)  } <\omega_{1}$,
the interval Ia is restricted by the inequality $q<q_{c4}\equiv2\sqrt{2m\mu}$, while the interval
Ib is not restricted, see Fig. \ref{fig:AngPoints}. In this case we show results
only for window Ib. When $\max_q  \omega_{2}^{\left(
ph\right)  }  >\omega_{1}$, interval Ib disappears at $q_{c1}<q<q_{c2}$,
so interval Ia is preferable
in this figure. Below, poles of the GPF propagator are used for an analytic simulation of spectral functions.
The selection of appropriate poles is automatically determined by an interval between
two neighboring angular points for the frequency argument $\omega$ of the spectral function.
This completely resolves the question of a choice of a preferable interval for poles
to reproduce the spectral function for a given $\omega$.
} collective excitations
at $1/\left(  k_{F}a_{s}\right)=0$ and two different temperatures
to illustrate the changes taking place when approaching $T_c$.
Following the convention of Ref.~\cite{Klimin2019}, we call here
the ``first branch'' ($z_{I}^{(1)}$) the one whose sound
velocity continuously evolves to the velocity of first sound at $T=0$.
In the BCS regime ($1/k_F a\lesssim 0.155$, see the discussion in section
VI. B. 1. in \cite{Klimin2019}), this first branch always has a larger
sound velocity, \textit{i.e.} $\omega_{I}^{(1)}>\omega_{I}^{(2)}$ when $q\to 0$ and for all $T$.
At $1/k_F a > 0.155$,  it is rather the imaginary part which distinguishes the two modes:
the first branch always has (at all $T$ and this time also all $q$) a lower imaginary
part: $\gamma_{I}^{(1)}<\gamma_{I}^{(2)}$.

For larger momenta, the phononic branches behave strongly
nonlinearly and non-monotonically. For $T=0.9T_c$, the second eigenfrequency
passes a maximum and then goes down. The first eigenfrequency
continues to increase after its linear start
eventually crossing the pair-breaking continuum edge $\omega_{1}$.
When the branch reaches the pair-breaking continuum edge, its damping diminishes (as in the low temperature
case \cite{Kurkjian2016}), while the other solution becomes overdamped.
Although the phononic branch no longer fullfills the piecewise rule
when it is above $\omega_1$, this penetration in the continuum
suggests the presence of an observable resonance inside the pair-breaking continuum,
whose lower tails extends to the phononic intervals Ia and Ib. We will show below
that this resonance is nothing else than the developing pairing mode
$z^{(T_c)}$.

Remarkably, the high-$q$ behavior of the first and second poles is switched
when temperature increases from $T=0.9T_{c}$ to $T=0.95T_{c}$.
While at $T=0.9T_{c}$ the frequency of the second pole
is always below that of the first pole, at $T=0.95T_{c}$ its frequency is lower than that
of fhe first pole at small momentum, but larger at higher $q$,
penetrating into the pair-breaking continuum.
This demonstrates an avoided crossing of complex poles when varying
temperature. We remind that an analogous phenomenon was observed in  Ref. \cite{Klimin2019}
for complex sound velocities. The two complex poles behave as repulsive particles in the complex $z$
plane with one of the parameters $(q,T,1/a_s)$ playing the role of time.
In between the left and right panels of Fig.~\ref{fig:AB1}, there exists
a specific temperature $T_{ex}$ corresponding to a precise
\textquotedblleft head-on collision\textquotedblright\ of the two poles in an
exceptional point $\left(  q_{ex},T_{ex}\right)  $, where crossing and
anticrossing cannot be distinguished. At $1/a_{s}=0$ this
exceptional point lies at $q_{ex}\approx0.556\sqrt{2m\mu}$ and
$T_{ex}\approx0.933T_{c}$.
When varying the interaction strength, both $T_{ex}$ and
$q_{ex}$ increase towards the BCS regime (with $T_{ex}$
in particular tending to $T_c$ when $1/k_F a\to-\infty$). $q_{\rm ex}$
eventually vanishes at $1/k_Fa\approx0.155$ when the sound velocites show an exact crossing \cite{Klimin2019}.
Then, at $1/k_Fa>0.155$ the situation is simpler:
the second root always has a larger imaginary part (at all $q$ and all temperature),
and never penetrates the pair-breaking continuum.
In this regime, the second root thus has little physical significance
(even when its sound velocity is above that of the first root,
as this happens above a crossing temperature).

We note the interesting mathematical properties
of the exceptional point $(q_{ex}, T_{ex})$ where the two roots
are equal and thus indistinguishable. It constitutes a second order
branching point of the functions $(q,T)\mapsto z_{I}^{(1)}(q,T)$ and
$(q,T)\mapsto  z_{I}^{(2)}(q,T)$: close contours
with a winding number of $\pm1$ around $(q_{ex}, T_{ex})$
exchange $z_{I}^{(1)}$ and $z_{I}^{(2)}$. Anticipating on the discussion of section \ref{visibility},
we also note a divergence of the residues of both poles at the exceptional point.


\subsection{Collective excitations provided by intervals inside the pair-breaking continuum}
\label{continuum}

\subsubsection{Pair-breaking branch}

\begin{figure}%
\centering
\includegraphics[
height=4.8282in,
width=3.5181in
]%
{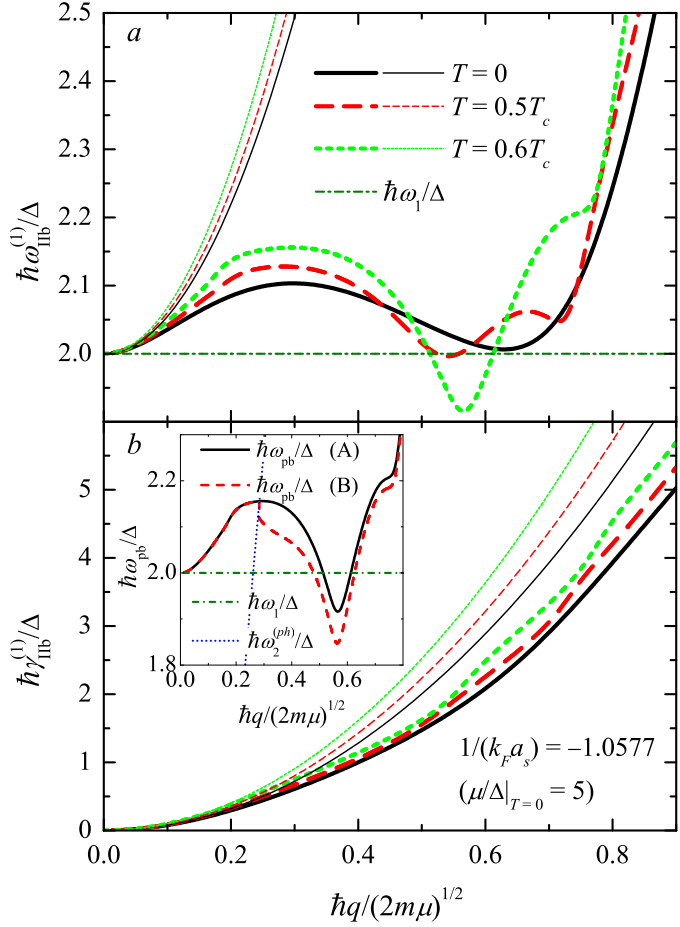}%
\caption{Frequency $\omega_{\rm IIb}^{\left(  1\right)  }\left(  q\right)  $
(panel \emph{a}) and damping factor $\gamma_{\rm IIb}^{\left(  1\right)  }\left(
q\right)  $ (panel \emph{b}) of the pair-breaking collective excitations as a
function of momentum at different temperatures for $1/k_{F}a_{s}%
\approx-1.0577$, which corresponds to $\left.  \mu/\Delta\right\vert _{T=0}%
=5$. Thin curves show the low-$q$ expansion, and heavy curves are results of
the full calculation. The dot-dashed line indicates the pair-breaking
continuum edge. \emph{Inset}: the frequency at $T=0.6T_{c}$ calculated using
different windows for the analytic continuation. The dotted curve is the
angular-point frequency $\omega_{2}^{\left(  ph\right)  }$.}%
\label{fig:QdepBCS}%
\end{figure}

In this paragraph, a special attention is paid to the finite-temperature
behavior of the pair-breaking collective excitations, treated previously at
$T=0$ in Ref. \cite{Kurkjian2019}. According to the classification of
intervals for the analytic continuation in Fig. \ref{fig:AngPoints}, there
exist two windows relevant for the pair-breaking branch of collective
excitations: IIa and IIb. Since the solutions in either continuation
are never too far apart\footnote{We show both solutions (\emph{a,~b}) only once in
the inset of Fig. \ref{fig:QdepBCS}, to demonstrate that they are indeed close
to each other within the linewidth determined by the damping factor. In other
figures, we avoid duplication of physically equivalent solutions.}
(their frequency separation $|\omega_{\rm IIb}-\omega_{\rm IIa}|$
in particular is much lower than their damping rate, which makes them indistinguishable
on the real axis), we always use window IIb which provides solutions
at all momenta (whereas window IIa is limited to $q_{c1}<q<q_{c2}$).

In Fig.~\ref{fig:QdepBCS}, the frequency and the damping factor are plotted as
functions of momentum for different temperatures with the same value of the
inverse scattering length $1/k_{F}a_{s}\approx-1.0577$ as above in Fig.
\ref{fig:AngPoints}. This value of the inverse scattering length is in
the BCS regime.
For comparison, the results of the small-momentum expansion developed in Ref.
\cite{Scirep} have been added to the figures (short-dashed and dot-dashed
curves). As can be seen from Fig. \ref{fig:QdepBCS}, the quadratic series
expansion is close to the result of the full calculation for $q\lesssim
0.02\sqrt{2m\mu}$. It cannot capture however a non-monotonic behavior of the
dispersion of pair-breaking modes at larger $q$. Conversely, the damping factor $\gamma_{\mathbf{q}}$
monotonically rises when increasing $q$. For $\hbar q\gtrsim\sqrt{2m\mu}$, the
pair-breaking mode frequency shifts to higher energy in the continuum, becoming strongly damped.
This behavior is qualitatively common for the zero and
non-zero temperatures.

For relatively low temperatures, the momentum dependence of the pair-breaking
mode frequency and damping is qualitatively close to that at $T=0$ reported in
Ref. \cite{Kurkjian2019}. For higher temperatures, however, a qualitative
difference appears. At sufficiently high temperatures, as we can see from Fig.
\ref{fig:QdepBCS} (\emph{a}), the mode
frequency exhibits oscillations just before moving to the overdamped regime.
Those oscillations are not visible in the spectral functions, and their
magnitude is relatively small with respect to damping. Hence they are not an
observable phenomenon, rather a mathematical peculiarity of the analytic continuation.

\begin{figure}[tbh]%
\centering
\includegraphics[
height=4.6665in,
width=3.5354in
]%
{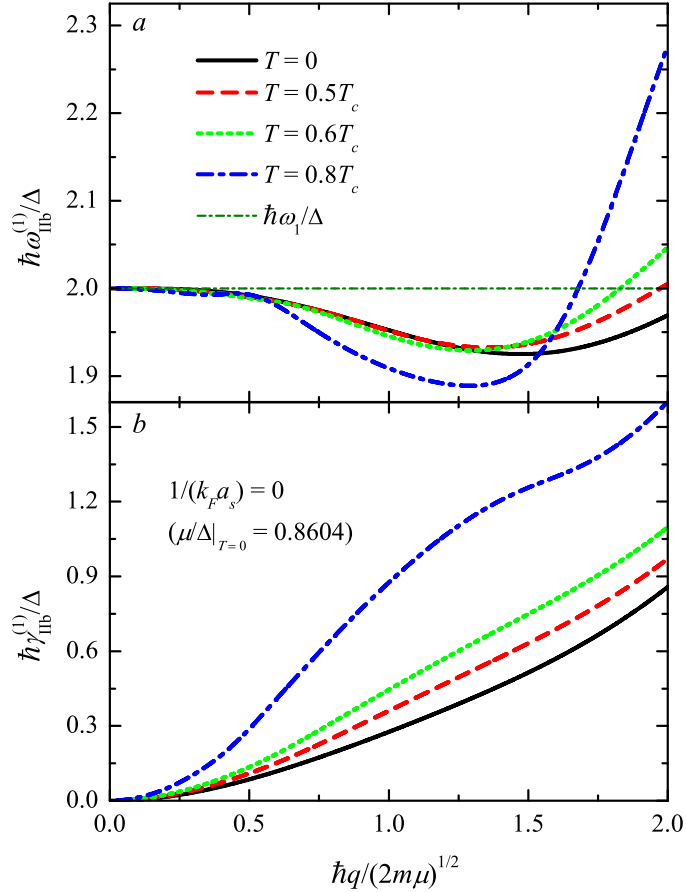}%
\caption{Frequency $\omega_{\rm IIb}^{\left(  1\right)  }\left(  q\right)  $
(panel \emph{a}) and damping factor $\gamma_{\rm IIb}^{\left(  1\right)  }\left(
q\right)  $ (panel \emph{b}) of the pair-breaking collective excitations as a
function of momentum at different temperatures for $1/k_{F}a_{s}=0$, which
corresponds to $\left.  \mu/\Delta\right\vert _{T=0}\approx0.8604$.
The dot-dashed line indicates the pair-breaking continuum edge.}%
\label{fig:QdepU}%
\end{figure}

At unitarity, the quadratic start of pair-breaking mode eigenfrequency is slower, as
can be seen from Fig. \ref{fig:QdepU}. There (and for
stronger couplings), the sign of the dispersion is negative at low $q$, so that the
frequency goes to the \textquotedblleft forbidden\textquotedblright\ area
$\omega_{\rm II}^{(1)}<\omega_{1}$, when using the analytic continuations through
both window IIa and IIb. At large $q$, the dispersion becomes non-monotonic.
For sufficiently high momentum, the eigenfrequencies can therefore
appear above the pair-breaking continuum, being however
substantially damped. As found in \cite{Kurkjian2019}, the damping of pair-breaking modes at unitarity is
smaller than in the BCS regime.

\begin{figure}[tbh]%
\centering
\includegraphics[
height=4.6873in,
width=3.4912in
]%
{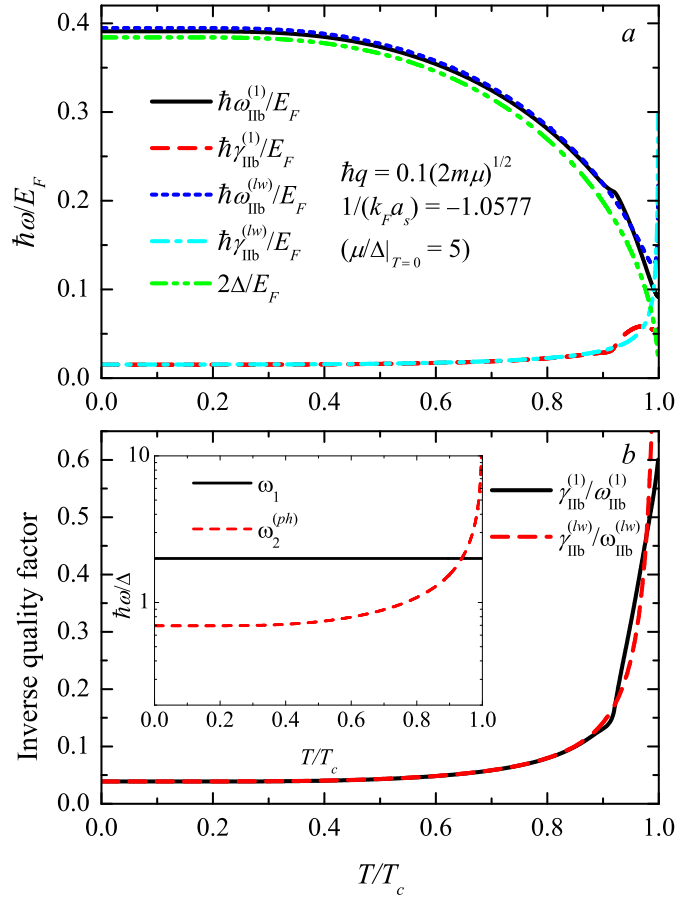}%
\caption{(\emph{a}) Temperature dependence of the frequency and damping of
pair-breaking collective excitations (in units of $E_F$) with momentum $q=0.1\sqrt{2m\left.
\mu\right\vert _{T=0}}$ for $1/k_{F}a_{s}\approx-1.0577$. Solid
and dashed curves show, respectively, the frequency and the damping factor as
functions of $T/T_{c}$. Dotted and dot-dashed curves represent the results of the small-$q$
expansion \cite{Scirep}. The dot-dot-dashed curve shows $2\Delta$.
(\emph{b}) The dimensionless inverse quality factor $\gamma_{\mathbf{q}}/\omega_{\mathbf{q}%
}$ calculated within the full calculation and the low-momentum expansion.
\emph{Inset}: temperature dependence of the boundary frequencies $\omega_{1}$
(solid curve) and $\omega_{2}^{\left(  ph\right)  }$ (dashed curve).}%
\label{fig:HMTQ01}%
\end{figure}

Fig. \ref{fig:HMTQ01} show the temperature dependence of the frequency and
damping of pair-breaking collective excitations for $1/k_{F}a_{s}%
\approx-1.0577$ at the particular value of momentum $q=0.1\sqrt{2m\left.
\mu\right\vert _{T=0}}$. Also $2\Delta$ has been plotted at the same
graph. In panel (\emph{b}) of the figure, we plot the inverse quality
factor $\gamma_{\mathbf{q}}/\omega_{\mathbf{q}}$ as a function of $T/T_{c}$.
The results of the full calculation within the present arbitrary-momentum
method are compared with the results of the small-momentum expansion
\cite{Scirep}. As can be seen from Fig. \ref{fig:HMTQ01}, the low-momentum
expansion agrees well with the full calculation at a relatively small momentum
$q=0.1\sqrt{2m\left.  \mu\right\vert _{T=0}}$ everywhere except in a
temperature range close to the transition temperature, where the
 long-wavelength expansion (limited to $q\ll 1/\xi_{\rm pair}$) is no longer valid
 for the selected value of $q$.
The small-momentum expansion exhibits a
divergence for both the frequency and the damping factor when $T$ tends to $T_c$.
On the contrary, the full finite-momentum calculation predicts finite values for the frequency and
the damping factor in the limit $T\rightarrow T_{c}$.%

In the inset to Fig. \ref{fig:HMTQ01}, we plot the particle-particle and
particle-hole angular-point frequencies. For $q=0.1\sqrt{2m\left.
\mu\right\vert _{T=0}}$, they cross each other at a temperature relatively
close to $T_{c}$. This explains the fast non-monotonic behavior of the
eigenfrequency and the damping factor at $T$ close to $T_{c}$ in Fig.
\ref{fig:HMTQ01} (\emph{a}) and the failure of the long-wave length expansion,
limited near $T_{c}$ to $\omega_{2}^{\left(  ph\right)  }<\omega_{1}$, that is
$q^{2}\ll\Delta^{2}/\mu$.

\begin{figure}[tbh]%
\centering
\includegraphics[
height=4.7323in,
width=3.4912in
]%
{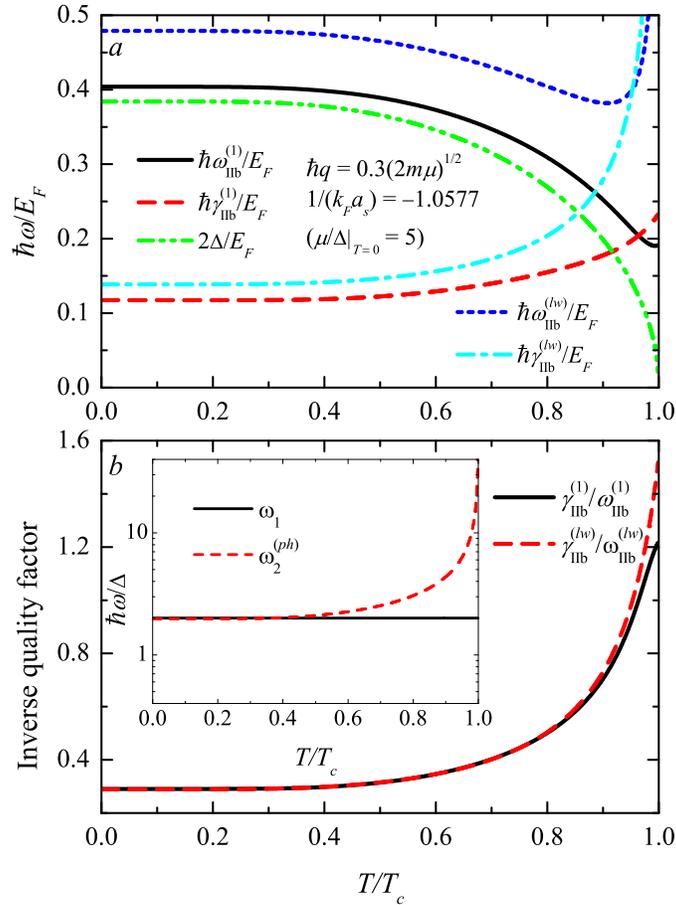}%
\caption{Temperature dependence of the frequency, damping and the inverse
quality factor of pair-breaking collective excitations with momentum
$q=0.3\sqrt{2m\left.  \mu\right\vert _{T=0}}$ for $1/k_{F}a_{s}\approx
-1.0577$. The notations are the same as in Fig. \ref{fig:HMTQ01}.}%
\label{fig:HMTQ03}%
\end{figure}

The temperature dependence of the eigenfrequency and the damping factor for a
higher momentum $q=0.3\sqrt{2m\left.  \mu\right\vert _{T=0}}$ is plotted in
Fig. \ref{fig:HMTQ03}. The difference between the results of the full-momentum
calculation and the small-$q$ expansion is here larger than in the case of
smaller momentum, but the temperature dependence remains qualitatively the
same although it is smoother.
The inset to Fig. \ref{fig:HMTQ03} shows plot the particle-particle and
particle-hole angular-point frequencies. For $q=0.3\sqrt{2m\left.
\mu\right\vert _{T=0}}$, as compared with the result shown in Fig.
\ref{fig:HMTQ01}, this crossing is relatively smooth and at a lower
temperature, so that we do not observe a fast change of frequencies and
damping factors in Fig. \ref{fig:HMTQ03}.

\subsubsection{Pole-doubling near $T_c$ and interplay with branches in windows III and IV}

\begin{figure}[ptbh]%
\centering
\includegraphics[
height=4.2289in,
width=5.9785in
]%
{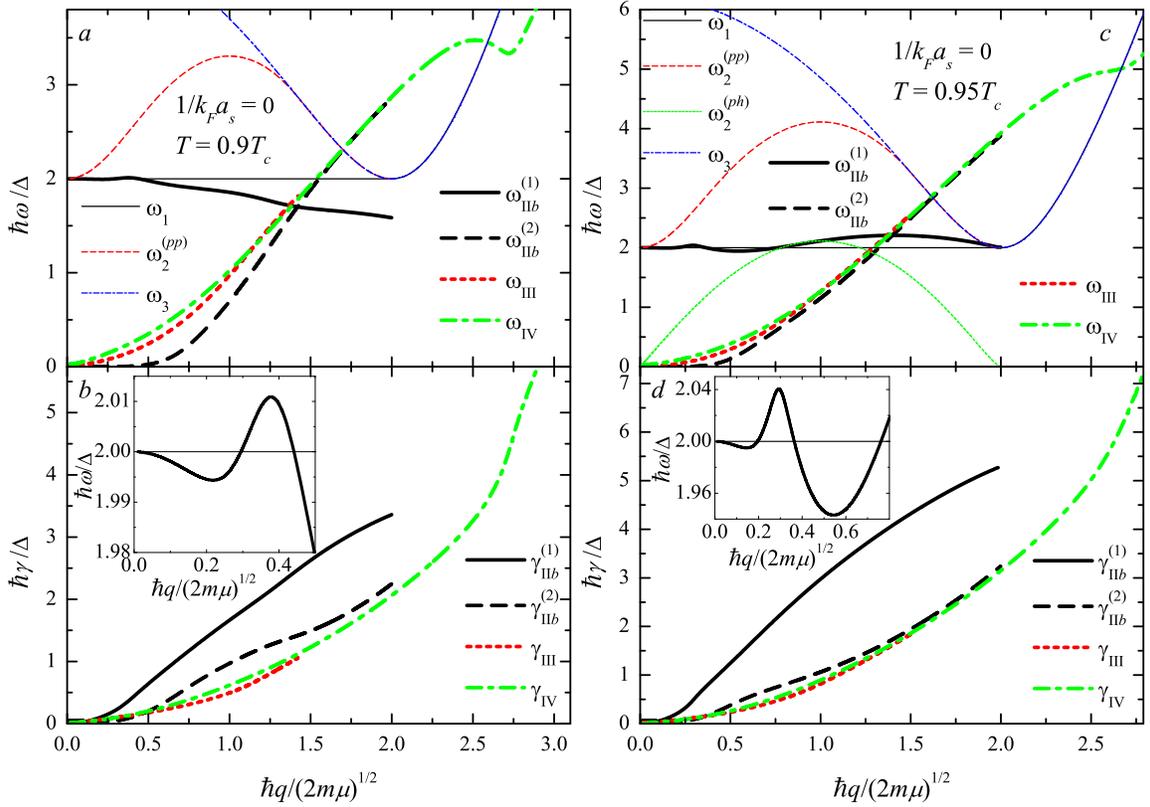}%
\caption{Momentum dependence of eigenfrequencies (\emph{a, c}) and damping
factors (\emph{b, d}) for collective excitations determined using the analytic
continuation through intervals positioned above the pair-breaking continuum
edge. Here, $T=0.9T_{c}$ (\emph{a, b)}, $T=0.95T_{c}$ (\emph{c, d}), with
$1/\left(  k_{F}a_{s}\right)  =0$. Black heavy solid and dashed curves represent the
solutions corresponding, respectively, to the first and second pole of the GPF
propagator resolved through window II (here window IIb is used). Dotted red and dot-dashed green
lines show the (unique) pole in respectively window III and IV. Thin curves show the
angular-point frequencies $\omega_{1}$ (the pair-breaking continuum edge),
$\omega_{2}^{\left(  pp\right)  }$ (the particle-particle angular point
frequency), $\omega_{2}^{\left(  ph\right)  }$ (the particle-hole angular
point frequency), and $\omega_{3}$. \emph{Insets}: the low-momentum part of
the pair-breaking mode frequency scaled for a better resolution.}%
\label{fig:H1}%
\end{figure}

In this subsection, we consider the parallel evolution of all collective modes
obtained using the analytic continuation through intervals IIa, IIb, III and IV. These intervals are
positioned above the pair-breaking continuum such that the obtained solutions
take into account both particle-particle and particle-hole scattering processes.
In Fig. \ref{fig:H1}, we show their dispersion relations at unitarity and both $T/T_{c}=0.9$ and
$T/T_{c}=0.95$.

\begin{figure}[ptbh]%
\centering
\includegraphics[
width=3.5in
]%
{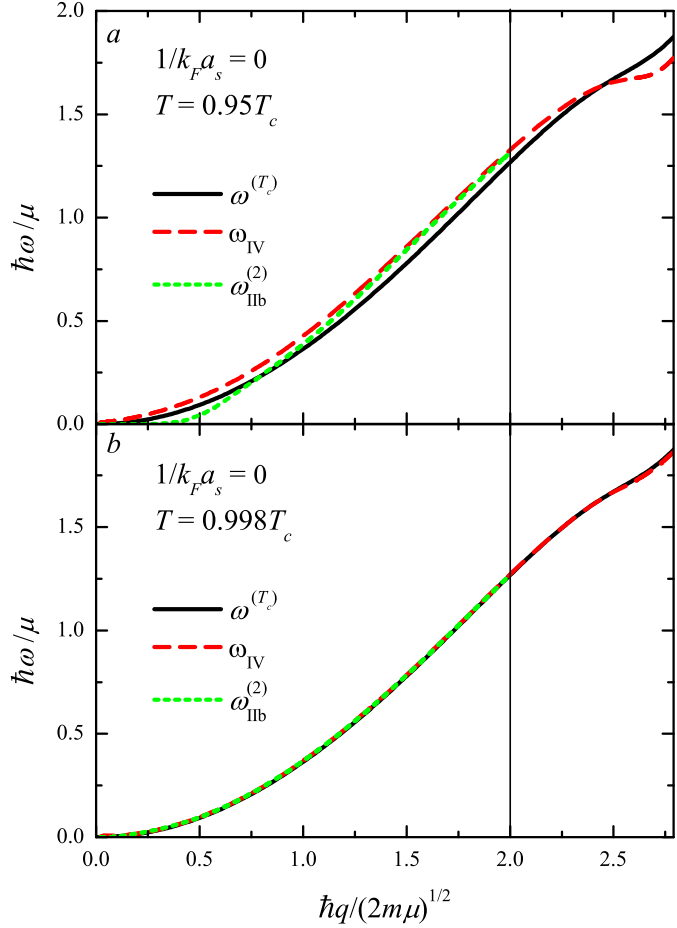}%
\caption{Momentum dispersion of the frequency of collective modes obtained using the
analytic continuation in frequency intervals IIb and IV at $1/\left(
k_{F}a_{s}\right)  =0$ with $T/T_{c}=0.95$ (\emph{a}) and $T/T_{c}=0.998$
(\emph{b}) compared with the eigenfrequency $\omega^{\left(T_{c}\right)}$
determined in Sec. \ref{Tc}. The vertical line shows the value of momentum $q = q_{c4}\equiv2\sqrt{2m\mu}$.}%
\label{figTc}%
\end{figure}

Remarkably, when $T$ is sufficiently close to $T_c$
a new pole  (shown by the black dashed curves on Fig.~\ref{fig:H1})
appears in the physical region ($\text{Re}z>0$ and $\operatorname{Im}z<0$)
of the window IIa and/or IIb. This is quite reminiscent of the pole-doubling already
observed (in section \ref{phononic} and Ref.~\cite{Klimin2019}) in the
phononic windows (Ia and Ib).
At low-$q$, the eigenfrequency of this pole tends to 0 (in contrast with the pair-breaking
\textquotedblleft Higgs\textquotedblright\ mode (black solid line), whose eigenfrequency
tends to $2\Delta$) and at $\hbar q\approx\sqrt{2m\mu}$ it
lies in the interval $[\omega_1,\omega_2]$ such that it fulfills the piecewise rule
(with a damping rate lower than that of the first pole).
This second pole thus seems to correspond
to a physically observable resonance when the momentum is sufficiently large.
In fact we show on Fig.~\ref{figTc}
that when $T\to T_c$, and for $1/\xi_{\rm pair}\ll q$, this pole 
tends asymptotically to the eigenenergy of the pairing collective mode $z_\qq^{(T_c)}$ found in section
\ref{Tc}. This demonstrates that the collective phenomenon we described near $T_c$
proceeds neither from the Anderson-Bogoliubov sound branch, nor from the pair-breaking ``Higgs''
branch which are characteristic of the low-temperature collective response.
The appearance of new poles in the analytic continuation, together with
the change in the dispersion relation, suggests
that we are dealing with a distinct physical phenomenon:
whereas the phononic and pair-breaking branches describe the collective
response of the pairs when they form a large fraction of the gas, the pairing
mode describes the response of an unpaired, or almost unpaired gas,
to externally driven pair formation.

For completeness, we also show on Fig.~\ref{fig:H1} the poles
found in window III (restricted to $q<q_{c3}$) and IV.
Both in frequency and damping $\omega_{\rm III}$ and $\omega_{\rm IV}$
are close  to the second pole of window II, $\omega_{\rm IIb}^{(2)}$
(that is, as long as $q$ is much larger than $1/\xi_{\rm pair}$).
Thus they also tend to $z_\qq^{(T_c)}$ when $T\to T_c$.
In fact, they represent the same physical resonance:
when $T= T_c$, the pair-breaking continuum no longer exhibits
the angular points $\omega_1$, $\omega_2$ and $\omega_3$
(except for $q>2\sqrt{2m\mu}$ where $\omega_3$ coincides with $\omega_0(q)$);
this means that the analytic continuations through windows II to IV
become simply equivalent.

At $q>2\sqrt{2m\mu}$ where only window IV remains,
we note that $\omega_{\rm IV}$ lies above $\omega_3$
(thus fulfilling the piecewise rule and providing
a sensible contribution to the pair field spectral function) in a wide range
of values of $q$. The eigenfrequency then shows a bump
when it crosses the pair-breaking continuum edge.
At very low $q\ll 1/\xi_{\rm pair}$, the solution $\omega_{\rm IV}$ is subsonic,
as one of the solutions
found by Ref.~\cite{Castin} at $T=0$. However it develops
the quadratic dispersion described in Sec. \ref{vicTc} in the intermediate regime
($1/\xi_{\rm pair}\ll q\ll k_F$).

\subsection{Visibility of the collective modes in spectral functions}
\label{visibility}

Now that we have extracted the collective mode spectrum
from the analytic continuations (both below (Sec.~\ref{phononic})
and above (Sec.~\ref{continuum}) the pair-breaking continuum),
we study the manifestations of this spectrum in the spectral
functions. Besides the pair response in the cartesian
basis introduced in Eq.~\eqref{sw1}, we study here
the modulus-modulus and phase-phase spectral functions:
\begin{align}
\chi_{aa}\left(  \mathbf{q},\omega\right)    & =\frac{1}{\pi}\operatorname{Im}%
\frac{Q_{1,1}\left(  \mathbf{q},\omega+\mathrm{i}0^{+}\right)  }%
{\det\mathbb{Q}\left(  \mathbf{q},\omega+\mathrm{i}0^{+}\right)  }%
,\label{sw2}\\
\chi_{pp}\left(  \mathbf{q},\omega\right)    & =\frac{1}{\pi}\operatorname{Im}%
\frac{Q_{2,2}\left(  \mathbf{q},\omega+\mathrm{i}0^{+}\right)  }%
{\det\mathbb{Q}\left(  \mathbf{q},\omega+\mathrm{i}0^{+}\right)  }.\label{sw3}%
\end{align}
We note that $\chi_{aa}$, $\chi_{pp}$ coincide when $T\to T_c$ at fixed $q$.
{Mathematically, this is because $Q_{11}=Q_{22}
=(M_{11}+M_{22})/2$ when $M_{12}$ can be neglected.}

\subsubsection{Residues of the complex poles}

\begin{figure}[ptbh]%
\centering
\includegraphics[
height=4.3777in,
width=6.0096in
]%
{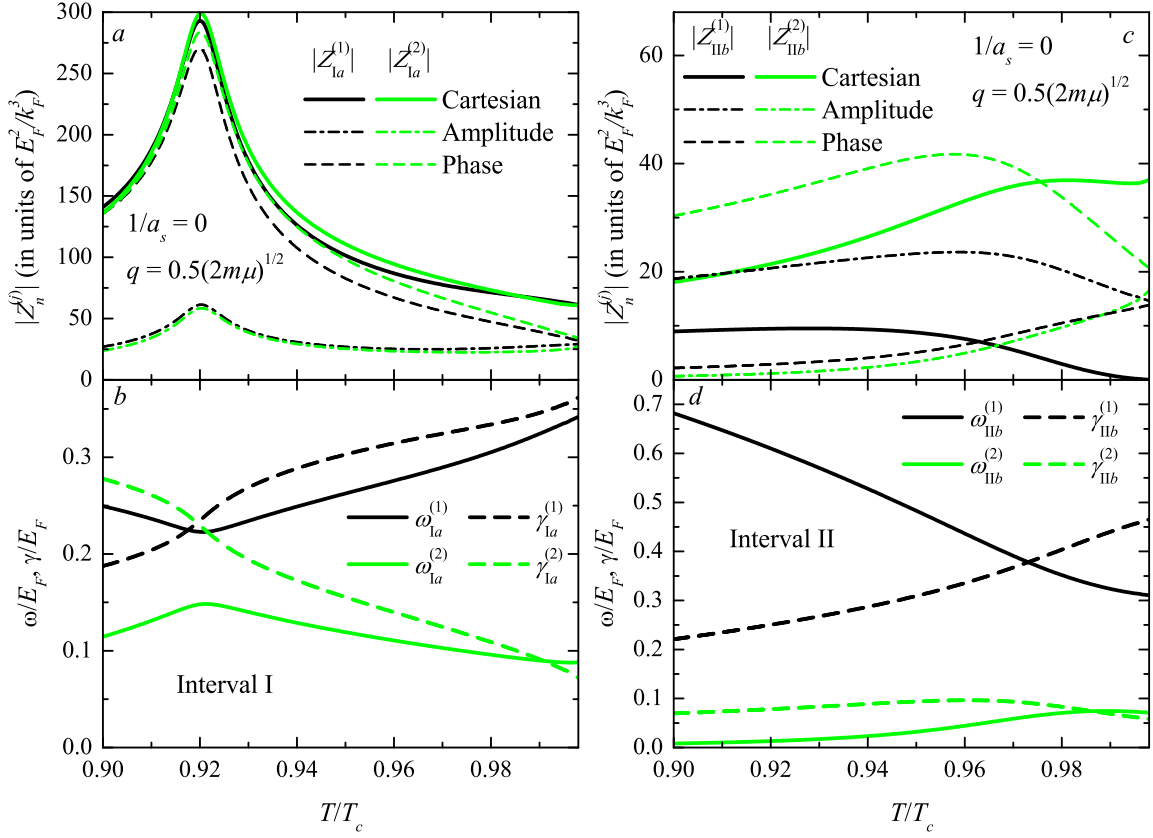}%
\caption{(\emph{a}, \emph{c}) Moduli of collective mode residues
in the cartesian channel (of spectral function $\chi$, \emph{solid curves}),
amplitude-amplitude channel (\emph{dot-dashed curves})  and phase-phase channel (\emph{dashed curves})
as functions of the relative temperature $T/T_{c}$ for the inverse scattering
length $1/a_{s}=0$ and the momentum $q=0.5\sqrt{2m\mu}$. (\emph{b}, \emph{d})
Frequencies (\emph{solid curves}) and damping factors (\emph{dashed curves})
for the same collective excitations. Panels (\emph{a}, \emph{b}) and (\emph{c},
\emph{d}) show collective excitations of window Ia (below the
continuum) and IIb (inside the continuum), respectively. The exceptional point
here is $(q_{ex},T_{ex})\approx(0.556\sqrt{2m\mu}, 0.933T_{c})$.
The results for the first and second pole are shown, respectively, by black
and green (light grey in grayscale) curves.}%
\label{fig:Residues}%
\end{figure}

\paragraph{Phononic branches}
We first analyse the residues $Z_{\mathbf{q}}$
of different complex poles in $\chi_{aa}$ and $\chi_{pp}$.
Fig. \ref{fig:Residues}.\emph{a} shows the residues for the phonon-like branches
$z_{\rm Ia}^{\left(  1\right)  }$ and $z_{\rm Ia}^{\left(  2\right)
}$ as functions of the relative temperature $T/T_{c}$ for a fixed
momentum $q=0.5\sqrt{2m\mu}$.
The value of $q$ is chosen because it lies slightly below the
exceptional-point value $q_{ex}$ where the two poles undergo a
head-on collision. Panel \ref{fig:Residues}.\emph{a} then illustrates the
behavior of the residues near the exceptional point.
The absolute values $|Z_{\rm Ia}^{(1)}|$ and $|Z_{\rm Ia}^{(2)}|$ of the residues
of the two phononic modes show a resonant increase near $T_{ex}$ (we checked
numerically their divergence when $\left(  q,T\right)  \rightarrow\left(  q_{ex},T_{ex}\right)$).
However, because residues are complex and the phase of the residues are
opposite at resonance, this does not result in a resonant enhancement
of the spectral function when the temperature passes $T_{ex}$, as will be
shown below in Fig. \ref{fig:SWfuns}.
The residues of the first and second modes remain
close (in absolute value) in the whole considered interval of temperatures
(both in the phase-phase and modulus-modulus channels). What determines the domination
of $z_{\rm Ia}^{(2)}$ over $z_{\rm Ia}^{(1)}$ in the spectral functions at temperatures
$T_{c}-T \ll T_c-T_{\rm ex}$ (and vice-versa the domination of $z_{\rm Ia}^{(1)}$
over $z_{\rm Ia}^{(2)}$ at $T\ll T_{\rm ex}$) is rather the crossing of the damping factors near $T_{\rm ex}$
(see panel \ref{fig:Residues}.\emph{b}).

The comparison of the phase
and modulus residues also shows a clear domination
of the phase channel (recall that in the long wavelength
limit \cite{Klimin2019} one has $|Z_{\rm Ia, phase-phase}|\gg|Z_{\rm Ia, modulus-modulus}|$)
except in the vicinity of $T_c$ where they converge to the same value, as expected.

\paragraph{Poles of window II}

Panel \emph{c} represents the residues of the two modes obtained using
the analytic continuation through the window IIb: the pair-breaking
\textquotedblleft Higgs \textquotedblright\ mode with the complex pole
$z_{\rm IIb}^{\left(  1\right)  }$ and the second mode  $z_{\rm IIb}^{\left(  2\right)  }$ of the same
window, which tends to the pairing mode $z_\qq^{(T_c)}$ when $T\to T_c$. The eigenfrequencies and
damping factors for these poles are shown in Fig. \ref{fig:Residues}
\emph{d}. This further illustrates how the pairing mode of Sec.~\ref{Tc}
replaces the pair-breaking mode as $T$ approaches $T_c$:
in the cartesian spectral function $\chi$, the residue
of $z_{\rm IIb}^{\left(  1\right)  }$ tends to 0 in the cartesian
spectral function $\chi$. Even in the modulus-modulus channel the residue
of $z_{\rm IIb}^{\left(  2\right)  }$ eventually
dominates over that of $z_{\rm IIb}^{\left(  1\right)  }$.
This effect adds up to a purely spectral effect (see panel \emph{d}):
the pair-breaking mode $z_{\rm IIb}^{\left(  1\right)  }$ becomes overdamped near $T_{c}$ while the other
mode $z_{\rm IIb}^{\left(  2\right)  }$ exhibits the opposite trend, it is
overdamped at lower temperatures, and its damping decreases when $T$
approaches $T_{c}$.


\subsubsection{Analytic simulations of the spectral functions}
We now wish to measure the amount of information on the spectral
functions which is contained in the spectrum and residues found
in the analytic continuation. For this, we define (as in Ref. \cite{Klimin2019})
``analytic simulations'' of the spectral functions using the poles $z_{\mathbf{q}}$
and residues $Z_{\mathbf{q}}$ in each of the 5 continuation windows:%
\begin{align}
\chi_{\mathrm{eff}}\left(  \mathbf{q},\omega\right)   &  =\sum_{n=1}^5\sum_{j_{n}%
}\delta\chi_{n}^{\left(  j_{n}\right)  }\left(  \mathbf{q},\omega\right)
,\label{hieff}\\
\delta\chi_{n}^{\left(  j_{n}\right)  }\left(  \mathbf{q},\omega\right)   &
=\frac{1}{\pi}\operatorname{Im}\left(  \frac{Z_{\mathbf{q},n}^{\left(
j_{n}\right)  }}{\omega-z_{\mathbf{q},n}^{\left(  j_{n}\right)  }}\right)
\Theta\left(  \Omega_{n-1}<\omega<\Omega_{n}\right)  , \label{hieff2}%
\end{align}
where the index $n$ in a partial contribution $\delta\chi_{n}^{\left(
j_{n}\right)  }$ indicates the interval used for the analytic continuation, and $j_n$
labels the complex poles of the GPF propagator continued through this interval.
Here, $\left\{  \Omega_{n}\right\}  $ are the angular-point frequencies as described above,
completed by $\Omega_{0}\equiv0$ and $\Omega_{5}\rightarrow\infty$ and sorted
in the ascending order as follows: $\Omega_{1}\equiv\min\left(  \omega
_{1},\omega_{2}^{\left(  ph\right)  }\right)  $, $\Omega_{2}\equiv\max\left(
\omega_{1},\omega_{2}^{\left(  ph\right)  }\right)  $, $\Omega_{3}\equiv
\omega_{2}^{\left(  pp\right)  }$, and $\Omega_{4}\equiv\omega_{3}$.
Particularly for $q>q_{c3}$, the interval between $\omega_{2}^{\left(
pp\right)  }$ and $\omega_{3}$ shrinks to zero and does not contribute to
$\chi_{\mathrm{eff}}$. The Heaviside step function $\Theta\left(  \Omega
_{n-1}<\omega<\Omega_{n}\right)  $ is used on the same reasoning as in Ref.
\cite{Klimin2019}: poles from an analytic continuation are relevant for the
spectral function only in the interval through which
the analytic continuation passed. It should be noted that the step
function means the piecewise rule for the argument $\omega$ of the spectral
function but \emph{not} a piecewise rule for complex poles
$z_{\mathbf{q},n}^{\left(  j_{n}\right)  }$. For a fixed
contribution $\chi_{n}^{\left(  j_{n}\right)  }\left(  \mathbf{q}%
,\omega\right)  $, real parts of relevant poles may move
beyond the interval $\Omega_{n-1}<\omega<\Omega_{n}$, as discussed in the caption of
Fig.~\ref{fig:schema}.

\begin{figure}[ptbh]%
\centering
\includegraphics[
height=4.8793in,
width=6.4904in
]%
{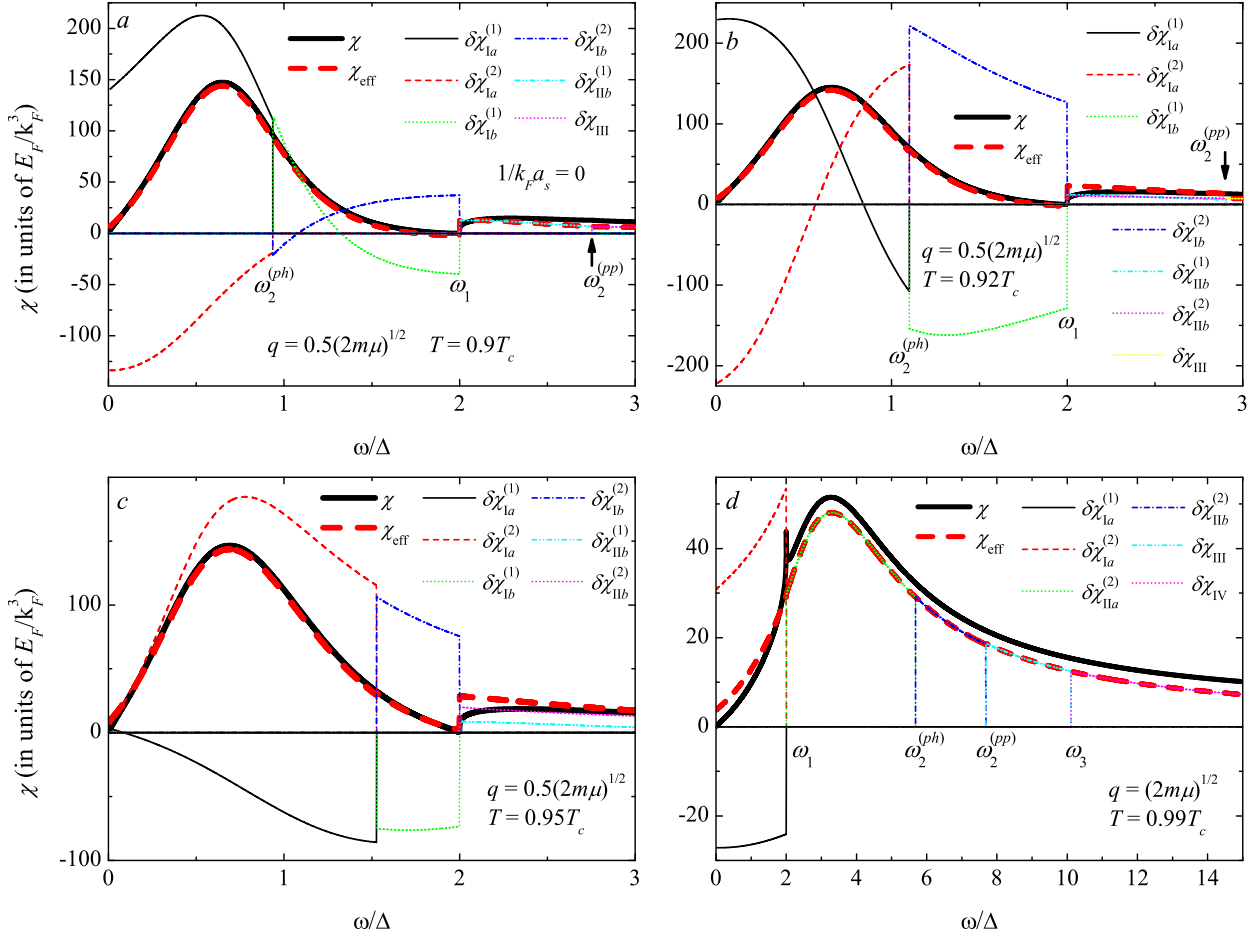}%
\caption{Spectral function $\chi\left(  \mathbf{q},\omega\right)$
(\emph{heavy black solid curve}), its analytic simulation $\chi_{\mathrm{eff}%
}\left(  \mathbf{q},\omega\right)  $ (\emph{heavy red dashed curve}) and
partial contributions of different complex poles (\emph{thin curves}) as
functions of the frequency for the inverse scattering length $1/a_{s}=0$ and
the momentum $q=0.5\sqrt{2m\mu}$ (\emph{a, b, c}) and $q=\sqrt{2m\mu}$
(\emph{d}) at the temperatures $T=0.9T_{c}$ (\emph{a}), $T=0.92T_{c}$
(\emph{b}), $T=0.95T_{c}$ (\emph{c}), and $T=0.99T_{c}$ (\emph{d}).}%
\label{fig:SWfuns}%
\end{figure}

In Fig. \ref{fig:SWfuns}.\emph{a, b, c}, the spectral function
$\chi\left(  \mathbf{q},\omega\right)  $ and its analytic simulation
$\chi_{\mathrm{eff}}\left(  \mathbf{q},\omega\right)  $ are shown for the same
momentum $q=0.5\sqrt{2m\mu}$ as chosen in Fig. \ref{fig:Residues}, for three
temperatures: below, near and above the crossing-point temperature $T_{\rm ex}$ for damping
factors. Also, partial contributions of different poles to $\chi
_{\mathrm{eff}}$ are plotted. Fig. \ref{fig:SWfuns}.\emph{d} for
$T=0.99T_{c}$ and a larger wavevector ($q=\sqrt{2m\mu}$) shows the spectral
function and its analytic simulation when temperature moves closer to
$T_{c}$ and the momentum is sufficiently large so that the resonance peak
lies in the pair-breaking continuum. We note that since
the residues are complex, the partial contributions to $\chi_{\rm eff}$
are not everywhere positive.

In all four panels of \eqref{fig:SWfuns}, we observe that
the total analytic simulation is remarkably close to
the spectral function $\chi$. This indicates that the
poles found in the analytic continuation are a good
summary of the shape of the spectral function.
When the frequency $\omega$ passes a bound between
two neighboring intervals for the analytic continuation, the retained partial
contributions abruptly change due to the Heaviside function in \eqref{hieff}.
In some cases, this leads to a well expressed discontinuity of $\chi_{\rm eff}$
at a ``sharp'' angular point, as on panels \ref{fig:SWfuns} \emph{a},\emph{b},\emph{c}
at $\omega=\omega_1$. In other cases, the discontinuity of $\chi_{\rm eff}$ is much smaller than
the average value of the function, such that the discontinuity at a ``soft'' angular point
is hardly resolvable by eye. This happens either because two adjacent
intervals have almost the same poles in their analytic continuation,
and hence almost equal partial contributions
(the case of panel \ref{fig:SWfuns} \emph{d} at $\omega_2^{ph}$, $\omega_2^{pp}$ and $\omega_3$),
or, more subtly, because two partial contributions add up to almost
the same value of $\chi_{\rm eff}$, despite having an important discontinuity at the angular point
(the case of panel \ref{fig:SWfuns} \emph{b} at $\omega_2^{ph}$).
This distinction between the ``sharp'' angular point ($\omega_1$) and ``soft'' ones
($\omega_2^{ph}$, $\omega_2^{pp}$ and $\omega_3$) has a physical origin:
at the soft angular points only the configuration of resonant
wavevectors changes, not the damping mechanism itself. On the contrary, $\omega_1$
separates regions where the damping channel by emission of broken pairs is either opened or closed.
This being said, we note that the sharpness of $\omega_1$ decreases when $T\to T_c$
at fixed $q$ (see panel \ref{fig:SWfuns} \emph{d}). This is expected:
at $T\geq T_c$ the quasiparticle-quasiparticle continuum is no longer distinguishable
from the rest of the particle-hole continuum.

Having a comparable residue, the two phononic poles have overall comparable contributions.
At the resonance of residues (Fig. \ref{fig:SWfuns}~\emph{b})
the two poles participate in the peak of $\chi_{\mathrm{eff}%
}$ almost equally. When $T\to T_c$, the broadening of the partial contribution of $z_{\rm Ia}^{(1)}$
 makes its contribution near the peak of $\chi_{\rm eff}$ comparatively smaller.
It should be noted that the resonance of residues is not
manifested in the temperature behavior of the total spectral function.
Moreover, the change of a dominant partial contribution can hardly be
distinguished in the total response, and can be extracted only using the
analytic simulation.

When $T$ approaches $T_c$ (Fig. \ref{fig:SWfuns}~\emph{d}), the maximum of the pair field
response enters the pair-breaking continuum. In windows IIa and IIb,
the main peak almost entirely proceeds from $\delta\chi
_{\rm II}^{\left(  2\right)  }$, the contribution of the first pole (the pair-breaking ``Higgs
mode'') being negligible. The peak extends almost without discontinuity or angular points
onto window III and IV. This does not surprises us since $z_{\rm IIa}^{(2)}$, $z_{\rm IIb}$ and
$z_{\rm III}$ and $z_{\rm IV}$ all tend asymptotically to each other and to the $T_{c}$ pairing mode
$z^{\left(  T_{c}\right)  }$ analyzed in Sec. \ref{Tc} (recall Fig.~\ref{figTc}). Near $T_c$,
the angular points separating those 4 intervals tend to disappear, which makes
the analytic continuation through the 4 windows nearly equivalent.
Although the discussion is purely formal since the intervals are equivalent,
we note that the width of windows IIb and III tends to 0 as $T\to T_c$
such that only intervals IIa and IV remain.

\begin{figure}[ptbh]%
\centering
\includegraphics[
height=5.2667in,
width=6.442in
]%
{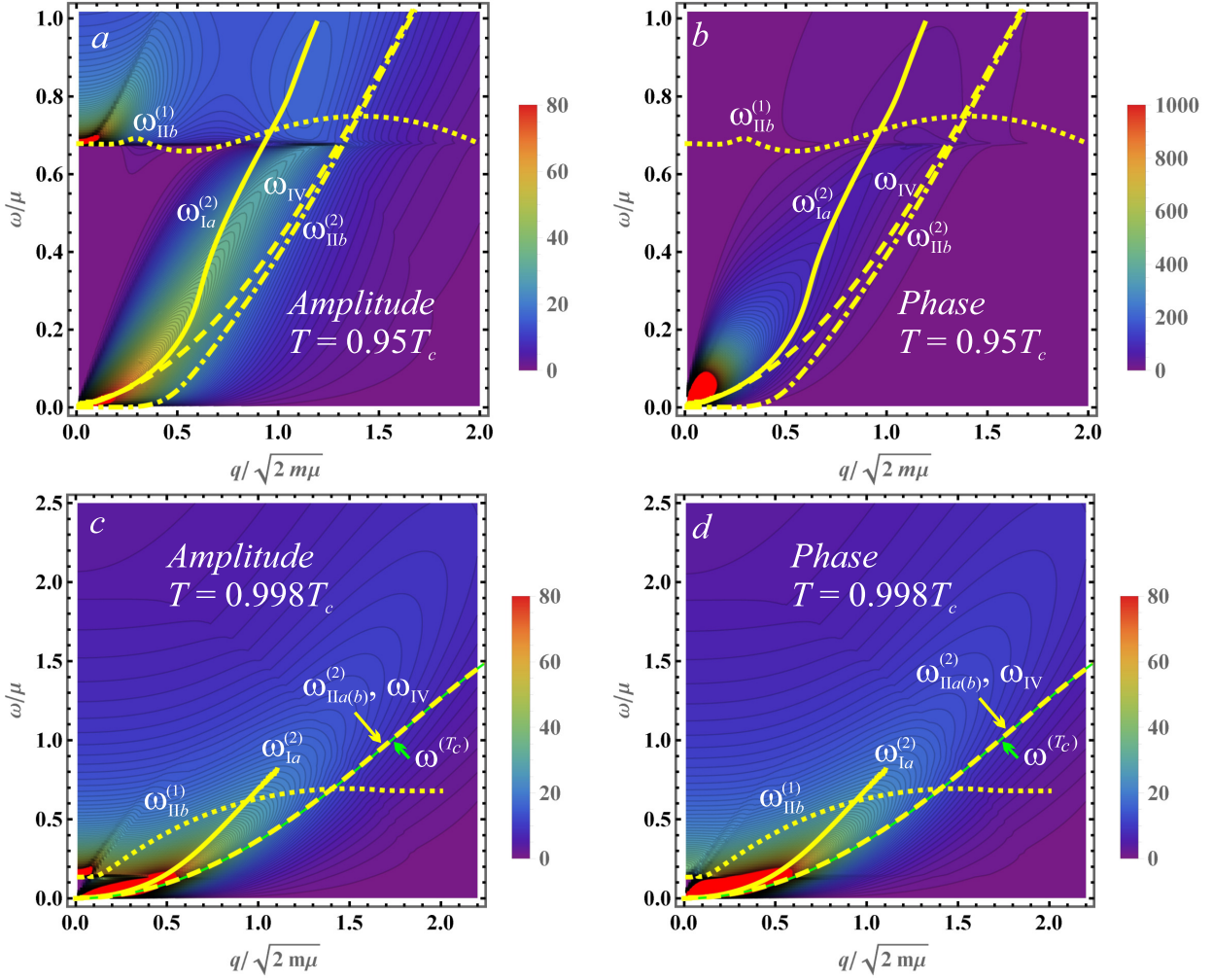}%
\caption{Contour plots of amplitude-amplitude (\emph{a,~c}) and phase-phase
(\emph{b,~d}) response functions (in units of $E_F/k_F^3$)
for $1/\left(  k_{F}a_{s}\right)  =0$
at $T/T_{c}=0.95$ (\emph{a,~b}) and $T/T_{c}=0.998$ (\emph{c,~d}). Yellow (white in grayscale) solid, dashed and dotted curves
show eigenfrequencies determined by poles of the GPF propagator explicitly indicated in the figure.
The solid green (light grey) curves show eigenfrequencies for the GPF propagator at $T=T_{c}$. Clipping
areas above the upper limits in color codes are shown by red (dark grey).}%
\label{Contours}%
\end{figure}

\subsubsection{Evolution of collective modes near $T_c$}

To better illustrate how the spectral function evolves from a phonon/Higgs mode
regime at low temperature, to a
regime dominated by the quadratic
pairing mode, we show on Fig.~\ref{Contours}
contour plots of the modulus-modulus and phase-phase spectral functions.
On top of the contour plots, we indicate selected eigenfrequencies from roots
of Eq.(\ref{det}). Here, only the roots which give the most significant contributions to the
spectral functions have been plotted\footnote{Here, the choice between
\textquotedblleft\emph{a}\textquotedblright\ and \textquotedblleft%
\emph{b}\textquotedblright\ windows needs an explanation. It depends on the
fact whether the analytic simulation $\chi_{\mathrm{eff}}$ using this window
is appropriate to reproduce the spectral function $\chi$. More clearly,
for $T=0.95T_{c}$, the \textquotedblleft Higgs\textquotedblright\ mode lies
above the angular-point frequency $\omega_{2}^{\left(  ph\right)  }$ in the
range of momenta where it is not overdamped. The sound-like
mode frequencies at the same temperature are mainly lower than $\omega
_{2}^{\left(  ph\right)  }$. Consequently, we plot here $\omega_{\mathrm{Ia}%
}^{\left(  2\right)  }$, $\omega_{\mathrm{IIb}}^{\left(  1\right)  }$, and
$\omega_{\mathrm{IIb}}^{\left(  2\right)  }$, choosing the interval
\textquotedblleft\emph{a}\textquotedblright\ for sound-like modes and
\textquotedblleft\emph{b}\textquotedblright\ for modes in the continuum. Also,
this explains a choice of representative intervals for the analytic
continuation in Fig. \ref{fig:Residues}. On the contrary, at $T=0.998T_{c}$,
the angular-point frequency $\omega_{2}^{\left(  ph\right)  }$ appears to be
higher than $\omega_{\mathrm{IIa}\left(
\mathrm{b}\right)  }^{\left(  2\right)  }$ in the range of $q$ where the
mode is not overdamped. Therefore the window \textquotedblleft\emph{a}%
\textquotedblright\ is relevant for Fig. \ref{Contours}
(\emph{c}, \emph{d}).}.
As $T$ approaches $T_{c}$, the region of energy-momentum where the influence
of the phononic-like modes $\omega_{\mathrm{Ia}}^{\left(  1,2\right)  }$ and
pair-breaking mode $\omega_{\mathrm{IIb}}^{\left(  1\right)  }$
shrinks to a small window $q\lesssim1/\xi_{\mathrm{pair}}$ and $\omega
\lesssim\Delta$, corresponding to the region where the existence of condensed pairs still
matters. Elsewhere, the spectral function is dominated by a
resonance well summarized by $z^{(T_{c})}\approx  z_{{\rm II}a}^{(2)}\approx  z_{{\rm II}b}^{(2)}\approx z_{\rm IV}$
(and the associated residue\footnote{We note that $T=0.998T_c$ and $q>0.5\sqrt{2m\mu}$,
the complex residue clearly shifts the peak of the resonance away from $\omega^{(T_{c})}$.}).
Fig. \ref{Contours} thus illustrates the reduction of the
phononic/pair-breaking regime when $T\rightarrow T_{c}$, and the corresponding growth of a
regime dominated by the pairing collective mode $z_{\mathbf{q}}^{(T_{c})}$.
Again we note that $\chi_{aa}$ and $\chi_{pp}$ coincide when
$T\rightarrow T_{c}$ at fixed $q$ as is clearly visible on the lower panels of Fig. \ref{Contours}.

\section{Conclusions}

We have investigated collective excitations in condensed Fermi gases in the
whole range of the BCS-BEC crossover for finite temperatures below $T_{c}$ and
beyond the small-momentum approximation. Eigenfrequencies and damping factors
for different branches of collective excitations are calculated within the
Gaussian pair fluctuation approach using a unified method of finding complex
poles of the analytically continued GPF propagator. The real and imaginary
parts of complex poles are calculated mutually consistently, beyond the
perturbation theory for damping. This makes it possible to consider collective
excitations also in cases when damping is not small.

At and near $T_c$, we showed that a quadratically-dispersed collective mode,
acting as a precursor of the phase transition, is observable in the response of the system to a
driving pairing field. This mode was predicted by Andrianov and Popov in the BCS limit \cite{Andrianov1976}
and appears in the dissipative time-dependent Ginzburg-Landau equation of Ref.~\cite{deMelo1993}.
We computed analytically its effective
mass and showed how it varies from purely imaginary values in the BCS limit to purely real
in the BEC limit. Away from $T_c$ we computed (to leading order in $|T-T_c|$)
the energy shift, which acts as a gap in the BEC regime and as an extra damping
rate in the BCS limit. Last, we explained how the resonance
disappears at large $q$ when it encounters the lower-edge of the pairing continuum.

We note that the drastic change in the dispersion of the collective modes
is predicted here within GPF theory, which approximates
the correlation length critical exponent to its BCS value $\nu=1/2$, in a quantitative disagreement
with the calculated value for the universality class of superfluid Fermi gases
$\nu\simeq0.62$  \cite{Wetterich2010,Dupuis2016}
and with the result of the conformal bootstrap calculation $\nu=0.6718(1)$ \cite{Chester2020}.
Integrating this correction to the study of collective modes
near the transition temperature
would certainly make the prediction more accurate.
This needs of course a calculation beyond the GPF approximation, which is not able to
study the critical regime quantitatively.
Our approach also assumes an infinite quasiparticle lifetime, which restricts
us for the study of collective modes to the collisionless regime. Extending our analysis
to the hydrodynamic regime, where excitations analogous to Carlson-Goldman modes
are expected, will be an important step forward. This requires a precise estimate
of the quasiparticle lifetime, which, in an ultracold Fermi gas weakly coupled to its
environment, should be limited by intrinsic processes such as quasiparticle collisions.

Away from the limits $T\rightarrow T_{c}$ and $q\rightarrow0$, our general
study allows us to track the evolution of different branches of collective
excitations, and to make clear genetic relations between them. Particularly,
we show that the collective mode near the transition temperature
is genetically distinct from both pair-breaking and phononic modes (whose
visibility domain shrinks to a small window $\omega\lesssim\Delta$ and $q\lesssim 1/\xi_{\rm pair}$
near $T_c$) as it is caused by the appearance of new poles in the analytic continuation.
At $0<T<T_{c}$, eigenfrequencies and damping factors exhibit a nontrivial momentum and
temperature dependence. Particularly, they can be non-monotonic as functions
of $q$. Moreover, different eigenfrequencies may cross each other and change
their relative significance when varying momentum and temperature.
The present study clarifies then some unexplored
questions in the theory of collective excitations in superfluid Fermi gases.
The applied method can be straightforwardly extended to more complicated
condensed fermionic systems, e.~g., multiband or color superfluids.

\appendix
\section{Fluctuation matrix}
\label{app:matrix}
We recall here the elements of the order-parameter
fluctuation matrix $\mathbb{Q}$ in modulus-phase basis (for the matrix $\mathbb{M}$,
in the cartesian basis see Eqs.~(10) and (11) in \cite{Klimin2019}).
\begin{align}
Q_{1,1}\left(  \mathbf{q},z\right)   &  =-\frac{1}{8\pi a_{s}}+\int%
\frac{d\mathbf{k}}{\left(  2\pi\right)  ^{3}}\left\{  \frac{1}{2k^{2}}%
+\frac{X\left( \beta E_{\mathbf{k}-\frac{\mathbf{q}}{2}}\right)  }{4E_{\mathbf{k}%
-\frac{\mathbf{q}}{2}}E_{\mathbf{k}+\frac{\mathbf{q}}{2}}}\right. \nonumber\\
&  \times\left[  \left(  \xi_{\mathbf{k}-\frac{\mathbf{q}}{2}}\xi
_{\mathbf{k}+\frac{\mathbf{q}}{2}}+E_{\mathbf{k}-\frac{\mathbf{q}}{2}%
}E_{\mathbf{k}+\frac{\mathbf{q}}{2}}-\Delta^{2}\right)  \left(  \frac
{1}{z-E_{\mathbf{k}-\frac{\mathbf{q}}{2}}-E_{\mathbf{k}+\frac{\mathbf{q}}{2}}%
}-\frac{1}{z+E_{\mathbf{k}-\frac{\mathbf{q}}{2}}+E_{\mathbf{k}+\frac
{\mathbf{q}}{2}}}\right)  \right. \nonumber\\
&  \left.  \left.  +\left(  \xi_{\mathbf{k}-\frac{\mathbf{q}}{2}}%
\xi_{\mathbf{k}+\frac{\mathbf{q}}{2}}-E_{\mathbf{k}-\frac{\mathbf{q}}{2}%
}E_{\mathbf{k}+\frac{\mathbf{q}}{2}}-\Delta^{2}\right)  \left(  \frac
{1}{z-E_{\mathbf{k}+\frac{\mathbf{q}}{2}}+E_{\mathbf{k}-\frac{\mathbf{q}}{2}}%
}-\frac{1}{z-E_{\mathbf{k}-\frac{\mathbf{q}}{2}}+E_{\mathbf{k}+\frac
{\mathbf{q}}{2}}}\right)  \right]  \right\}  , \label{Q11}%
\end{align}

\begin{align}
Q_{2,2}\left(  \mathbf{q},z\right)   &  =-\frac{1}{8\pi a_{s}}+\int%
\frac{d\mathbf{k}}{\left(  2\pi\right)  ^{3}}\left\{  \frac{1}{2k^{2}}%
+\frac{X\left( \beta E_{\mathbf{k}-\frac{\mathbf{q}}{2}}\right)  }{4E_{\mathbf{k}%
-\frac{\mathbf{q}}{2}}E_{\mathbf{k}+\frac{\mathbf{q}}{2}}}\right. \nonumber\\
&  \times\left[  \left(  \xi_{\mathbf{k}-\frac{\mathbf{q}}{2}}\xi
_{\mathbf{k}+\frac{\mathbf{q}}{2}}+E_{\mathbf{k}-\frac{\mathbf{q}}{2}%
}E_{\mathbf{k}+\frac{\mathbf{q}}{2}}+\Delta^{2}\right)  \left(  \frac
{1}{z-E_{\mathbf{k}-\frac{\mathbf{q}}{2}}-E_{\mathbf{k}+\frac{\mathbf{q}}{2}}%
}-\frac{1}{z+E_{\mathbf{k}-\frac{\mathbf{q}}{2}}+E_{\mathbf{k}+\frac
{\mathbf{q}}{2}}}\right)  \right. \nonumber\\
&  \left.  \left.  +\left(  \xi_{\mathbf{k}-\frac{\mathbf{q}}{2}}%
\xi_{\mathbf{k}+\frac{\mathbf{q}}{2}}-E_{\mathbf{k}-\frac{\mathbf{q}}{2}%
}E_{\mathbf{k}+\frac{\mathbf{q}}{2}}+\Delta^{2}\right)  \left(  \frac
{1}{z-E_{\mathbf{k}+\frac{\mathbf{q}}{2}}+E_{\mathbf{k}-\frac{\mathbf{q}}{2}}%
}-\frac{1}{z-E_{\mathbf{k}-\frac{\mathbf{q}}{2}}+E_{\mathbf{k}+\frac
{\mathbf{q}}{2}}}\right)  \right]  \right\}  , \label{Q22}%
\end{align}

\begin{align}
Q_{1,2}\left(  \mathbf{q},z\right)   &  =i\int\frac{d\mathbf{k}}{\left(
2\pi\right)  ^{3}}\frac{X\left(\beta  E_{\mathbf{k}-\frac{\mathbf{q}}{2}}\right)
}{4E_{\mathbf{k}-\frac{\mathbf{q}}{2}}E_{\mathbf{k}+\frac{\mathbf{q}}{2}}%
}\nonumber\\
&  \times\left[  \left(  \xi_{\mathbf{k}-\frac{\mathbf{q}}{2}}E_{\mathbf{k}%
+\frac{\mathbf{q}}{2}}+E_{\mathbf{k}-\frac{\mathbf{q}}{2}}\xi_{\mathbf{k}%
+\frac{\mathbf{q}}{2}}\right)  \left(  \frac{1}{z-E_{\mathbf{k}-\frac
{\mathbf{q}}{2}}-E_{\mathbf{k}+\frac{\mathbf{q}}{2}}}+\frac{1}{z+E_{\mathbf{k}%
-\frac{\mathbf{q}}{2}}+E_{\mathbf{k}+\frac{\mathbf{q}}{2}}}\right)  \right.
\nonumber\\
&  \left.  +\left(  \xi_{\mathbf{k}-\frac{\mathbf{q}}{2}}E_{\mathbf{k}%
+\frac{\mathbf{q}}{2}}-E_{\mathbf{k}-\frac{\mathbf{q}}{2}}\xi_{\mathbf{k}%
+\frac{\mathbf{q}}{2}}\right)  \left(  \frac{1}{z-E_{\mathbf{k}+\frac
{\mathbf{q}}{2}}+E_{\mathbf{k}-\frac{\mathbf{q}}{2}}}+\frac{1}{z-E_{\mathbf{k}%
-\frac{\mathbf{q}}{2}}+E_{\mathbf{k}+\frac{\mathbf{q}}{2}}}\right)  \right]  ,
\label{Q12}%
\end{align}%
\begin{equation}
Q_{2,1}\left(  \mathbf{q},z\right)  =-Q_{1,2}\left(  \mathbf{q},z\right)  .
\label{Q21}%
\end{equation}
where we use $\hbar=k_B=2m=1$ and $E_{\mathbf{k}}=\sqrt{\xi_{\mathbf{k}}^{2}+\Delta^{2}}$ is the BCS
excitation energy, $\xi_{\mathbf{k}}=k^{2}-\mu$ is the free-fermion energy,
and $X$ is the function%
\begin{equation}
X\left(  t\right)  =\tanh\left(  \frac{t}{2}\right)  , \label{X}%
\end{equation}
These matrix elements coincide with those introduced in Refs. \cite{Engelbrecht}.

\section{Details on the long wavelength calculation near $T_c$}
\label{details}

We give here more details on the calculation of the collective
mode spectrum leading to expression \eqref{zqaboveTc} above $T_c$ and
\eqref{APeq} below $T_c$. Above $T_c$, we add and subtract
the sum $\sum_\kk {X(\beta \xi_\kk)}/{2\xi_\kk}$ to $M_{11}$, leading to:
\be
M_{11} (z,\qq,\beta<\beta_c)=I_1(z,\qq,\beta)+I_0(\beta)
\label{M11aboveTc2}
\ee
with
\bea
I_1(z,\qq,\beta)&=&\sum_{\kk}\frac{X(\beta\xi_+)+X(\beta\xi_-)}{2(z-\xi_+-\xi_-)}+\sum_\kk \frac{X(\beta \xi_\kk)}{2\xi_\kk}\\
I_0(\beta) &=& \sum_\kk \frac{X(\beta_c \xi_\kk)-X(\beta \xi_\kk)}{2\xi_\kk}=-\frac{\beta-\beta_c}{2}\sum_{\kk} X'(\beta_c \xi_\kk)+O(\beta-\beta_c)^2
\eea
Omitting terms of order $(\beta-\beta_c)q^2/2m$, one can then approximate $I_1$
by its value in $\beta=\beta_c$:
\be
I_1(\omega\pm\ii0^+,\qq,\beta_c)=C\frac{q^2}{2m}-D{'}\omega\mp\ii D{''} \omega
\ee

Below $T_c$, one should take into account the non-vanishing off-diagonal matrix element $M_{12}$ as well as the deviation
of the diagonal elements $M_{11}(\Delta)-M_{11}(0)$ due to the non-zero value of the gap. We compute the
latter quantity by setting $z=0$ before expanding for $q\to0$:
\begin{multline}
M_{11}(z=0,\qq,\Delta(T),T)-M_{11}(z=0,\qq,\Delta=0,T)\underset{q\to0}{=}\\\sum_{\kk}\bbcro{\frac{X(\beta\xi_\kk)}{2\xi_\kk}-\frac{X(\beta\epsilon_\kk)}{2\epsilon_\kk}}-\Delta^2 \sum_{\kk} \frac{\beta\epsilon_\kk X'(\beta\epsilon_\kk)-X(\beta\epsilon_\kk)}{4\epsilon_\kk^3}+O(q^2)\\
=-2I_0(\beta)+O(\beta-\beta_c)^2 \label{M11res}
\end{multline}
To recognize $-I_0$ in the first sum of the right-hand-side of \eqref{M11res}, we have used the gap equation in the form
\be
\sum_{\kk}\frac{X(\beta\epsilon_\kk)}{2\epsilon_\kk}=\sum_{\kk}\frac{X(\beta_c\xi_\kk)}{2\xi_\kk}
\label{gap}
\ee
For the second sum, we use the link between $\Delta^2$ and $\beta-\beta_c$ obtained by expanding \eqref{gap} for $\beta\to\beta_c$
with $\Delta^2\propto\beta-\beta_c$:
\be
\Delta^2 \sum_{\kk} \frac{\beta_c\xi X'(\beta_c\xi)-X(\beta_c\xi)}{4\xi^3} =I_0(\beta)+O(\beta-\beta_c)^2=(\beta-\beta_c)E+O(\beta-\beta_c)^2
\ee
We then approximate
\begin{multline}
M_{11,\downarrow}(z,\qq,\Delta,T)\simeq M_{11,\downarrow}(z,\qq,0,T)+M_{11}(0,\qq,\Delta,T)-M_{11}(0,\qq,0,T)\\
\simeq C\frac{q^2}{2m}-(D{'}+\ii D{''})z-(\beta-\beta_c)E(\beta_c\mu)
\label{M11approx}
\end{multline}
To compute $M_{22,\downarrow}$, we use the equality on the real axis $M_{22}(\omega+\ii0^+)=M_{11}(-\omega-\ii0^+)$
before doing the analytic continuation
\be
M_{22,\downarrow}(z,\qq,\Delta,T) \simeq C\frac{q^2}{2m}+(D{'}-\ii D{''})z-(\beta-\beta_c)E(\beta_c\mu)
\label{M22approx}
\ee
Finally, for the off-diagonal element $M_{12}$, the value in $z=0$, $q\to0$ suffices to leading order in $\beta-\beta_c$:
\be
M_{12}(0,\qq,\Delta,\beta)=-\Delta^2\sum_{\kk}\frac{X(\beta\epsilon)-\beta\epsilon X'(\beta\epsilon)}{4\epsilon^3}+O(q^2)=I_0(\beta)+O(\beta-\beta_c)^2
\ee

\begin{acknowledgments}
We acknowledge financial support from the Research
Foundation-Flanders (FWO-Vlaanderen) Grant No. G.0618.20.N, and from
the research council of the University of Antwerp.
\end{acknowledgments}

\end{document}